\documentclass[12pt]{article}
\usepackage{placeins}
\usepackage{amsmath,amssymb,amsfonts,amsthm,bm}    
\usepackage{lineno,hyperref}
\usepackage{algorithm2e}
\usepackage{graphicx}
\usepackage{pdfpages}
\usepackage[onehalfspacing]{setspace}
\usepackage[margin=1.5in]{geometry}
\usepackage{subcaption}
\usepackage{setspace}
\usepackage{placeins}
\usepackage{url}
\usepackage{xcolor}
\usepackage{mathtools}
\usepackage{titlesec}
\usepackage{natbib}

\newcommand{\blind}{1}

\begin{document}

	\def\spacingset#1{\renewcommand{\baselinestretch}%
		{#1}\small\normalsize} \spacingset{1}

	
	\if1\blind
	{
		\title{\bf Tempus Volat, Hora Fugit - A Survey of Tie-Oriented Dynamic Network Models in Discrete and Continuous Time}
		\author{Cornelius Fritz\thanks{\href{mailto:cornelius.fritz@stat.uni-muenchen.de}{cornelius.fritz@stat.uni-muenchen.de} }, Michael Lebacher and G\"oran Kauermann\hspace{.2cm}\\
			Department of Statistics, Ludwig-Maximilians-Universit\"at M\"unchen \\
			}
	\date{}
		\maketitle
	} \fi
	
	\if0\blind
	{
		\bigskip
		\bigskip
		\bigskip
		\begin{center}
			{\LARGE\bf Dynamic Networks}
		\end{center}
		\medskip
	} \fi
	
	\bigskip
	
\begin{abstract}
Given the growing number of available tools for modeling dynamic networks, the choice of a suitable model becomes central. The goal of this survey is to provide an overview of tie-oriented dynamic network models. The survey is focused on introducing binary network models with their corresponding assumptions, advantages, and shortfalls. The models are divided according to generating processes, operating in discrete and continuous time. First, we introduce the Temporal Exponential Random Graph Model (TERGM) and the Separable TERGM (STERGM), both being \newline time-discrete models. These models are then contrasted with continuous process models, focusing on the Relational Event Model (REM). We additionally show how the REM can handle time-clustered observations, i.e., continuous time data observed at discrete time points. Besides the discussion of theoretical properties and fitting procedures, we specifically focus on the application of the models on two networks that represent international arms transfers and email exchange. The data allow to demonstrate the applicability and interpretation of the network models.

\end{abstract}%
\newpage
\section{Introduction}
	The conceptualization of systems within a network framework has become popular within the last decades, see \citet{kolaczyk2009} for a broad overview. This is mostly because network models provide useful tools for describing complex dependence structures and are applicable to a wide variety of research fields. In the network approach, the mathematical structure of a graph is utilized to model network data. A graph is defined as a set of nodes and relational information  (ties) between them. Within this concept, nodes can represent individuals, countries or general entities, while ties are connections between those nodes. Dependent on the context, these connections can represent friendships in a school \citep{raabe2019}, transfers of goods between countries \citep{ward2013}, sexual relations between people \citep{bearman2004} or hyperlinks between websites \citep{leskovec2009} to name just a few. 
	Given a suitable data structure for the system of interest, the conceptualization as a network enables analyzing dependencies between ties. A central statistical model that allows this is the Exponential Random Graph Model (ERGM, \citealp{robins2001}). This model permits the inclusion of monadic, dyadic and hyperdyadic features within a regression-like framework.
	
	Although the model allows for an insightful investigation of \textsl{within-network} dependencies, most real-world systems are typically more complex. This is especially true if a temporal dimension is added, which is relevant, as most systems commonly described as networks evolve dynamically over time. It can even be argued that most static networks are \textsl{de facto} not static but snapshots of a dynamic process. A friendship network, e.g., typically evolves over time and influences like reciprocity often follow a natural chronological order. 

	Of course, this is not the first paper concerned with reviewing temporal network models. \citet{goldberg2008} wrote a general survey covering a wide range of models. The authors laid the foundation for further articles and postulated a soft division of statistical network models into latent space \citep{hoff2002} and $p_1$ models \citep{holland1981}, all originating in the Edös-Rényi-Gilbert random graph models \citep{erdos1959, gilbert1959}. \citet{kim2018} give a contemporary update on the field of dynamic models building on latent variables. \citet{snijders2005} discusses continuous time models and reframes the independence and reciprocity model as a Stochastic Actor oriented Model (SAOM, \citealp{snijders1996}). \citet{block2018} provide an in-depth comparison of the Temporal Exponential Random Graph Model (TERGM, \citealp{hanneke2010}) and the SAOM with special focus on the treatment of time. Further, the ERGM and SAOM for networks which are observed at single time points are contrasted by \citet{block2019}, deriving theoretical guidelines for model selection based on the differing mechanics implied by each model. 
	
	\begin{figure}[t!]\centering
		\includegraphics[trim={0cm 0cm 0cm 0cm},clip,width=\textwidth]{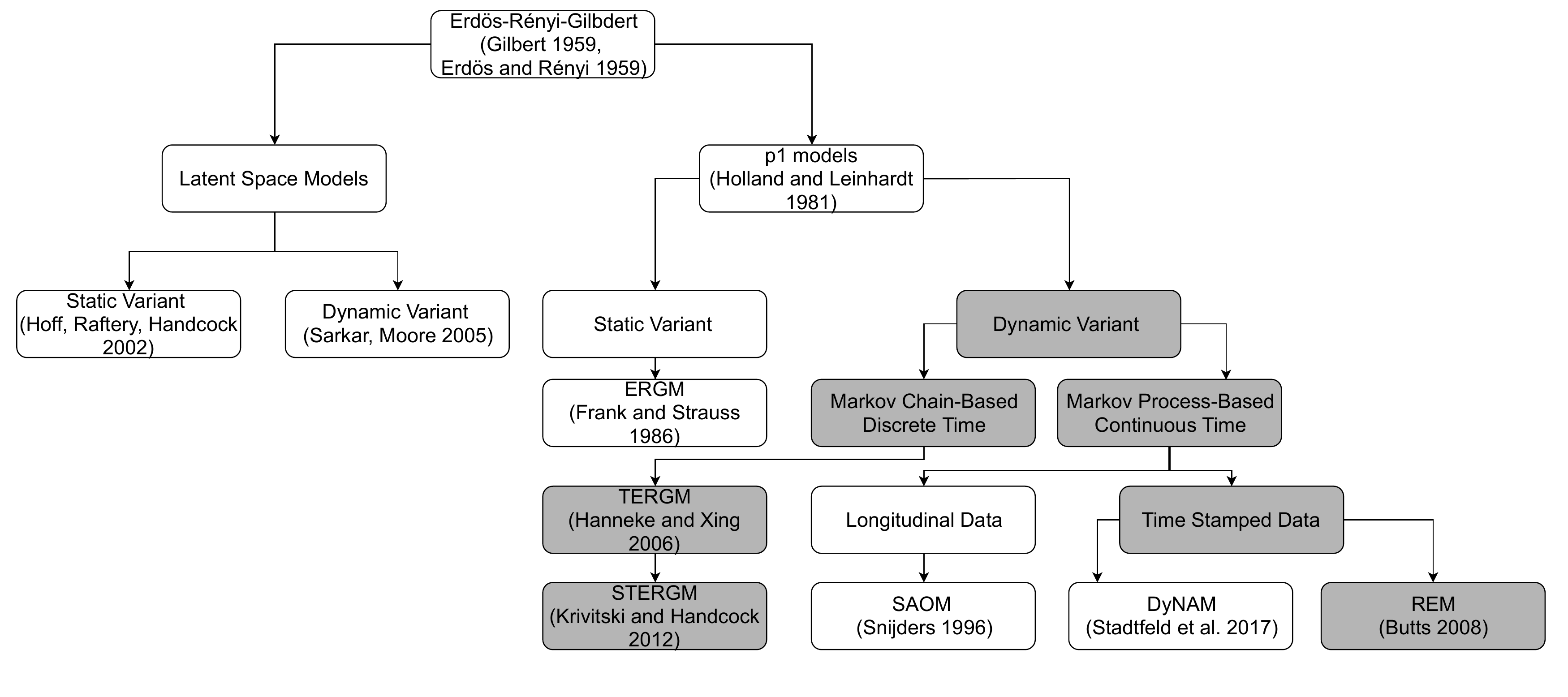}
		\caption{Tree diagram summarizing the dependencies between models originating in the Erdös-Rényi-Gilbert graph model, models situated in a box with a grey background are discussed in this article. This graph is an update of Figure 6.1 in \citet{goldberg2008}.}
		\label{fig:diag}
	\end{figure}
	
	In the context of this compendium of articles, the scope is to give an update on the dynamic variant of the second strand of models relating to $p_1$ models. We therefore extend the summarizing diagram of \citet{goldberg2008} as depicted in Figure \ref{fig:diag}. Generally, we divide temporal models into two sections, by differentiating between discrete and continuous time network models. In this endeavor, we emphasize reviewing tie-oriented models. Tie-oriented models are concerned with formulating a stochastic model for the existence of a tie contrasting the actor-oriented approach by \citet{snijders2002}, which specifies the model from the actors point of view \citep{block2018}.  The Dynamical Actor-Oriented model (DyNAM, \citealp{stadtfeld2017_4}) adopts this actor-oriented paradigm to event data. This type of model was formulated with a focus on social networks \citep{snijders1996}. Contrasting this, tie-oriented models can be viewed as more general, since they are also applicable to non-social networks.
	
	 Statistical models for time discrete data rely on an autoregressive structure and condition the state of the network at time point $t$ on previous states. This includes the TERGM and the Separable TERGM
	(STERGM, \citealp{krivitsky2014}). There exists a variety of recent applications of the TERGM. \citet{white2018covariation}  use a TERGM for modeling epidemic disease outcomes and \citet{blank2017political} investigate interstate conflicts. In \citet{HE2019443} Chinese patent trade networks are inspected and \citet{benton2017endogenous} use a TERGM for analyzing shareholder activism. Applications of STERGMs are given for example by \citet{stansfield2019sexual} that model sexual relationships and \citet{Broekel2019}  that study the network of research and development cooperation between German firms. 
	
	In case of time-continuous data, the model regards the network as a continuously evolving system. Although this evolution is not necessarily  observed in continuous time, the process is taken to be latent and explicitly models the evolution from the state of the network at time point $t-1$ to $t$ \citep{block2018}. In this paper we discuss the relational event model (REM, \citealp{butts2008}) for the analysis of event data. Applications of the REM are manifold and range from explaining the dynamics of health behavior sentiments via Twitter \citep{salathe2013}, inter-hospital patient transfers \citep{vu2017}, online learning platforms \citep{vu2015}, and animal behavior \citep{tranmer2015} to structures of project teams \citep{quintane2013}. Eventually, the REM is adapted to time-discrete observations of networks. That is, we observe the time-continuous developments of the network at discrete observation times only. Henceforth, we use the term \textsl{time-clustered} for this special data structure.
	
In reviewing dynamic network models, we assume a temporal first-order Markov dependency. To be more specific, this implies that the network at time point $t$ only depends on the previous observation of the network. This characteristic is widely used in the analysis of longitudinal networks \citep{hanneke2010,krivitsky2014} and the resulting conditional independence among states of the network facilitates the estimation with an arbitrary number of time points. In that respect, it suffices to only include two observational moments for illustrative purposes, since the interpretation and estimation with a longer series of networks is unchanged. Lastly, the comparison of the methods at hand in a clear-cut manner is hence enabled. 

	
	The paper is structured as follows. In Section \ref{sec:def} we give basic definitions that are used throughout the paper and present the two data examples that will be analyzed as illustrative examples. After that, Section \ref{sec:discrete} introduces time-discrete and Section \ref{sec:process} time-continuous network models. They are applied in Section \ref{Application} on two data sets and Section \ref{sec:discuss} concludes. Additional results relating to the applications can be found in the Supplementary Material.

	\nocite{sarkar2006}
	
	\section{Definitions and Data Description}\label{sec:def}
	\subsection{Definitions}
 This article regards directed binary networks, with ties representing directed relations between two nodes at a time point. The respective information can be represented in an adjacency matrix $Y_t=(Y_{ij,t})_{i,j=1,..,n} \in \mathcal{Y}$, where $\mathcal{Y} =\{Y: Y\in \lbrace 0,1 \rbrace ^{n \times n}\}$ represents the set of all possible networks with $n$ nodes. The entry $(i,j)$ of $Y_t$ is "1" if a tie is outgoing from node $i$ to  $j$ in year $t$ and "0" otherwise. Further, the discrete time points of the observations of $Y_t$ are denoted as $t = 1, \ldots, T$. We restrict our analysis to two time points in both exemplary networks, which suffices for comparison.
 Hence, we set $T=2$.  In many networks, including our running examples self-loops are meaningless. We therefore fix $Y_{ii,t} \equiv 0~ \forall~ i \in \lbrace 1, \ldots, n \rbrace$ throughout the article. Further, all sub-scripted temporal indices ($Y_t$) are assumed to take discrete and all indices in brackets ($Y(t)$) continuous values. The temporal indicator $t$ denotes the observation times of the network and to notationally differ this from time-continuous model we write $\tilde{t}$ for continuous time.  
 
 To sufficiently compare different models, we use two application cases. The first one represents the international trade of major weapons, which is given by discrete snapshots of networks that are yearly aggregated over time-continuous trade instances, i.e., the time-stamped information is not observed.
  Whereas the second application, a network of email traffic, comes in time-stamped format, that can be aggregated to discrete-time observations.
  
   	\begin{table}[t!] \centering 
   	\begin{tabular}{lccc|cc} 
   		\hline \hline
   		&  &\multicolumn{2}{c|}{Arms Trade Network}&\multicolumn{2}{c}{Email Network}\\\hline
   		Time&$t$ & 2016 & 2017 &Period $1$&Period $2$\\
   		Number of events& &$-$  & $-$ &$4\,957$&$2\,537$\\
   		Number of nodes &$n$& 180 & 180&88 &88 \\
   		Number of possible ties &$n(n-1)$& 32\,220 &  32\,220 &7\,656&7\,656 \\
   		Density & & 0.021 & 0.020 &0.123&0.087\\
   		Transitivity && 0.195 & 0.202& 0.407 & 0.345\\
   		Reciprocity && 0.081 & 0.083& 0.7 & 0.687\\
   		Repetition && $-$ & 0.641 & $-$& 0.574\\
   		\hline
   	\end{tabular} 
   	\caption{Descriptive statistics for the international arms trade network (left) and the European research institutions email correspondence (right). }
   	\label{tbl:descriptive} 
   \end{table}

		\subsection{Data Set 1: International Arms Trade}
		  \begin{figure}[t!]\centering
			\includegraphics[trim={0.6cm 3cm 0.6cm 3cm},clip,width=0.49\textwidth]{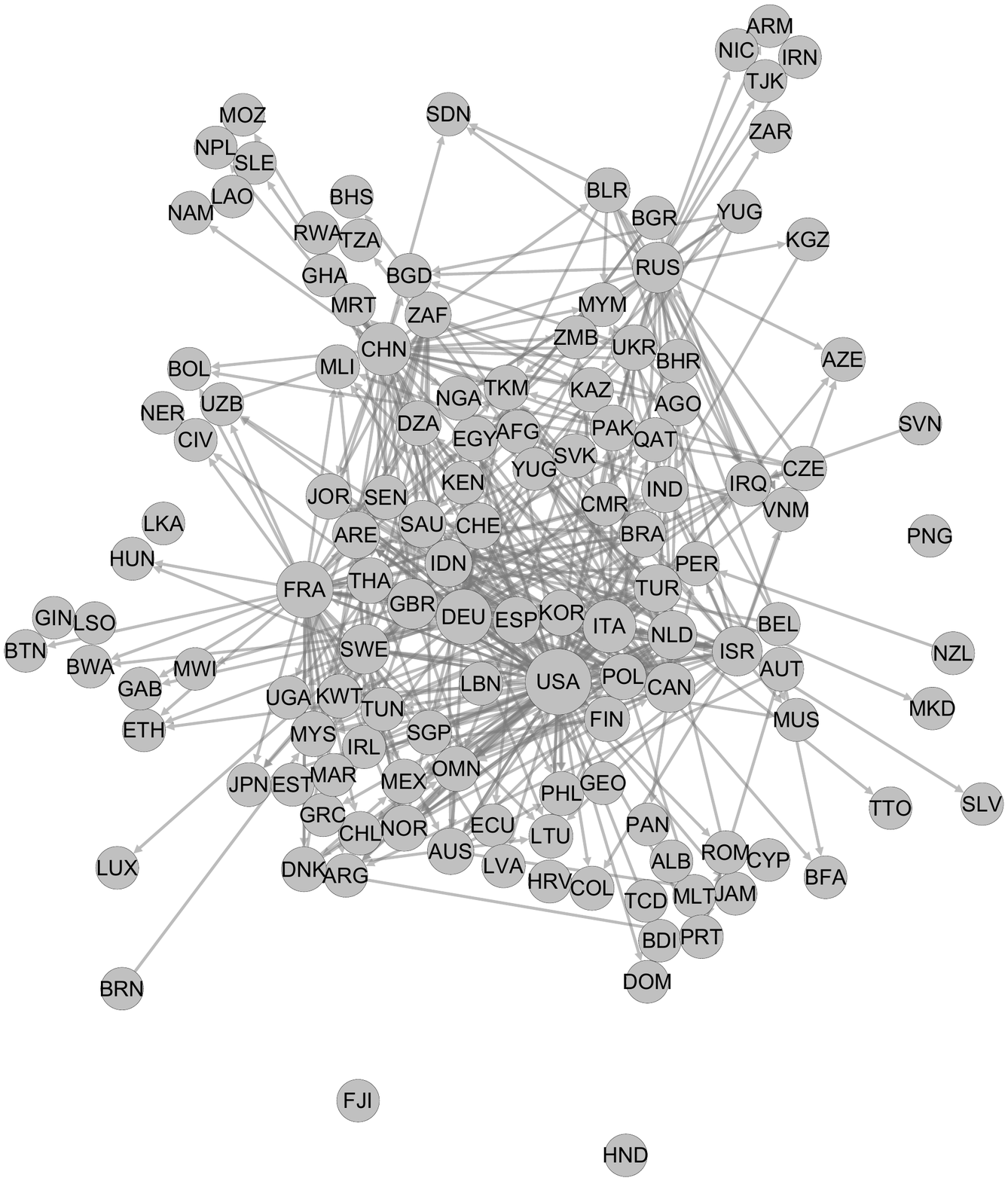}
			\includegraphics[trim={0.6cm 3cm 0.6cm 3cm},clip,width=0.49\textwidth]{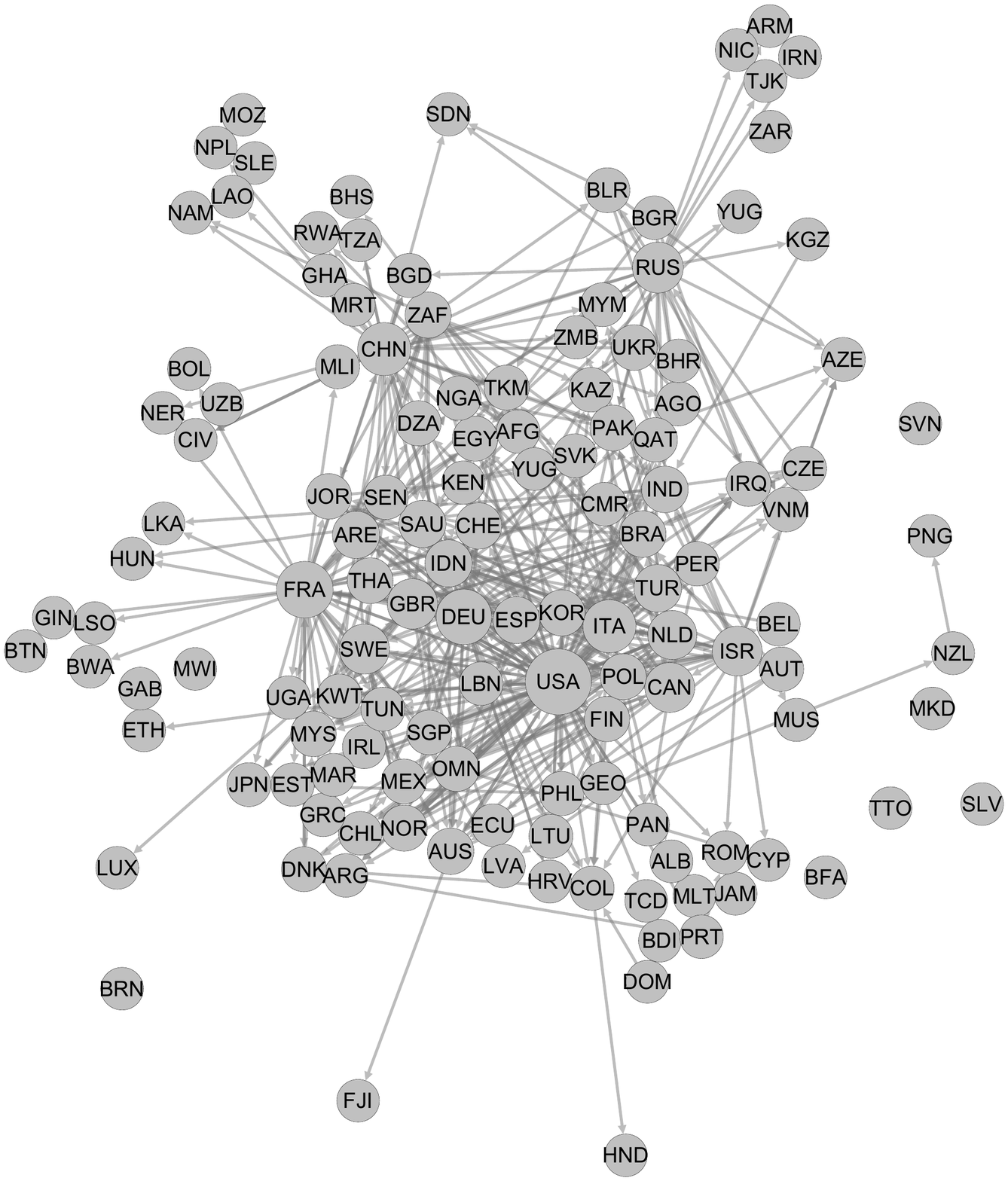}
			\caption{The International Arms Trade as a binary network in 2016 (left) and 2017 (right). Nodes that are isolated in both years are not depicted for clarity and the node size relates to the sum of the out- and in-degree. The labels of the nodes are the ISO3 codes of the respective countries.}
			\label{fig:networks}
		\end{figure}
		
The data on international arms trading for the years 2016 and 2017 are provided by the Stockholm International Peace Research Institute (\citealp{sipridata2017}). To be more specific, information on the exchange of major conventional weapons (MCW) together with the volume of each transfer is included. In order to have a binary network representation, we discretize the data and set edges $Y_{ij,t}$ to "1" if country $i$ sent arms to country $j$ in $t$.
		
		The left side of Table \ref{tbl:descriptive} gives some descriptive measures \citep{csardi2006} and Figure \ref{fig:networks} visualizes the arms trade network using the software \texttt{Gephi} (\citealp{ICWSM09154}). The density of a network is the proportion of realized edges out of all possible edges and is similar in both years, indicating the sparsity of the modeled network. Clustering can be expressed by the transitivity measure, providing the percentage of triangles out of all connected triplets. Reciprocity in a graph is the ratio of reciprocated ties and is similar in both years. As expressed by the high percentage of repeated ties, most countries seem to continue trading with the same partners. 
		
		Additionally, different kinds of exogenous covariates may be controlled for in statistical network models. In the given example we use the logarithmic Gross Domestic Product  (GDP) (\citealp{GDP2017}) as \textsl{monadic} covariates concerning the sender and receiver of weapons. We also include the absolute difference of the so called polity IV index (\citealp{Polity2016}), ranging from zero (no ideological distance) to 20 (highest ideological distance), as a \textsl{dyadic} exemplary covariate. These covariates are assumed to be non-stochastic and we denote them by $x_t$. See the Supplementary Material for a list of all included countries and their ISO3 code. 
		
		\subsection{Data Set 2: European Research Institution Email Correspondence}
		
The second network under study represents anonymized email exchange data between institution members of a department in a European research institution (see \citealp{paranjape2017motifs}, \citealp{mailcore}). In this data set, we observe events $\omega = (i,j,\tilde{t})$ that represent emails sent from department member $i$ to department member $j$ at a specific time point $\tilde{t}$. 

\begin{figure}[t!]\centering
	\includegraphics[trim={0cm 5cm 0cm 5cm},clip,width=0.48\textwidth]{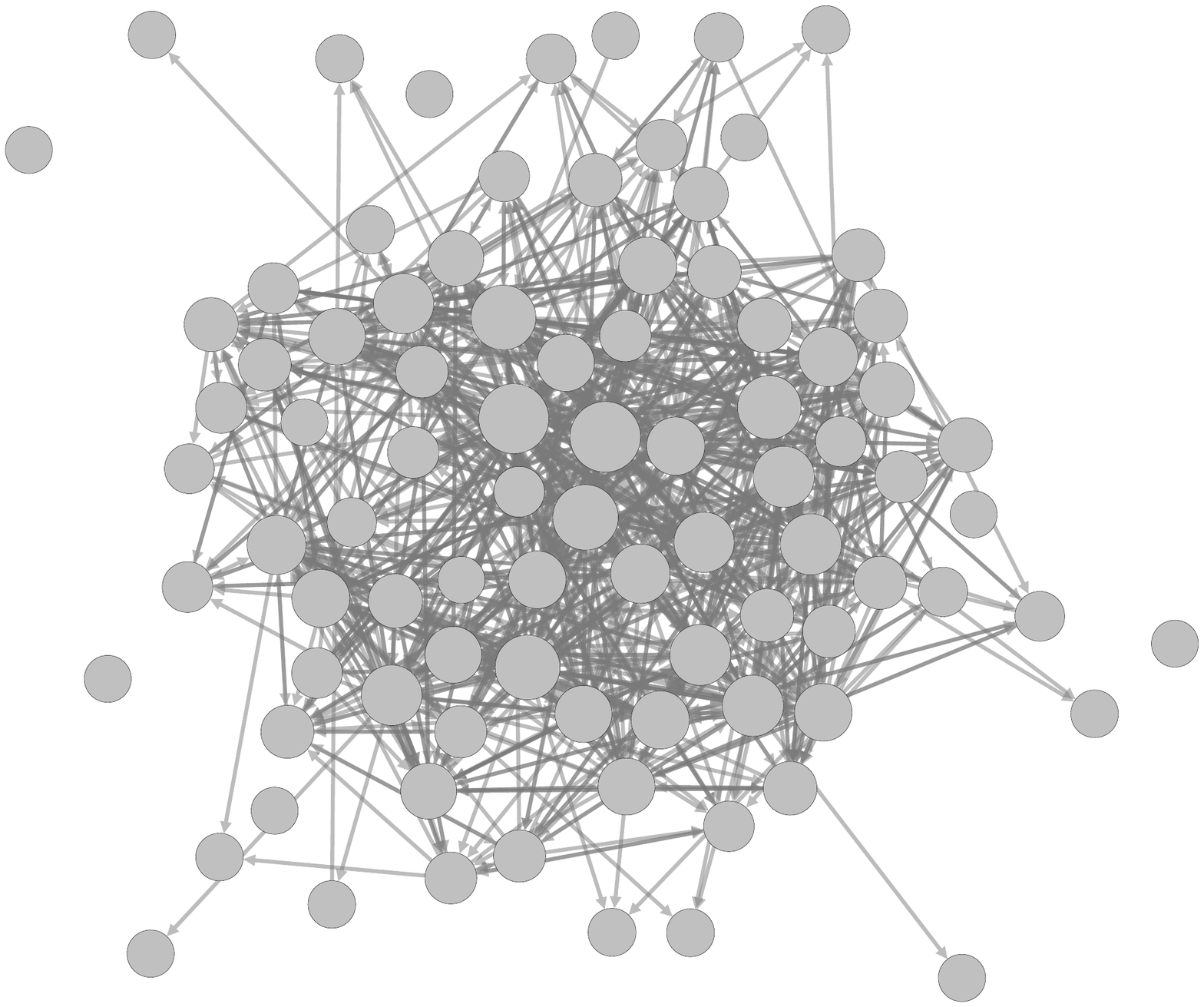}
	\includegraphics[trim={0cm 5cm 0cm 5cm},clip,width=0.48\textwidth]{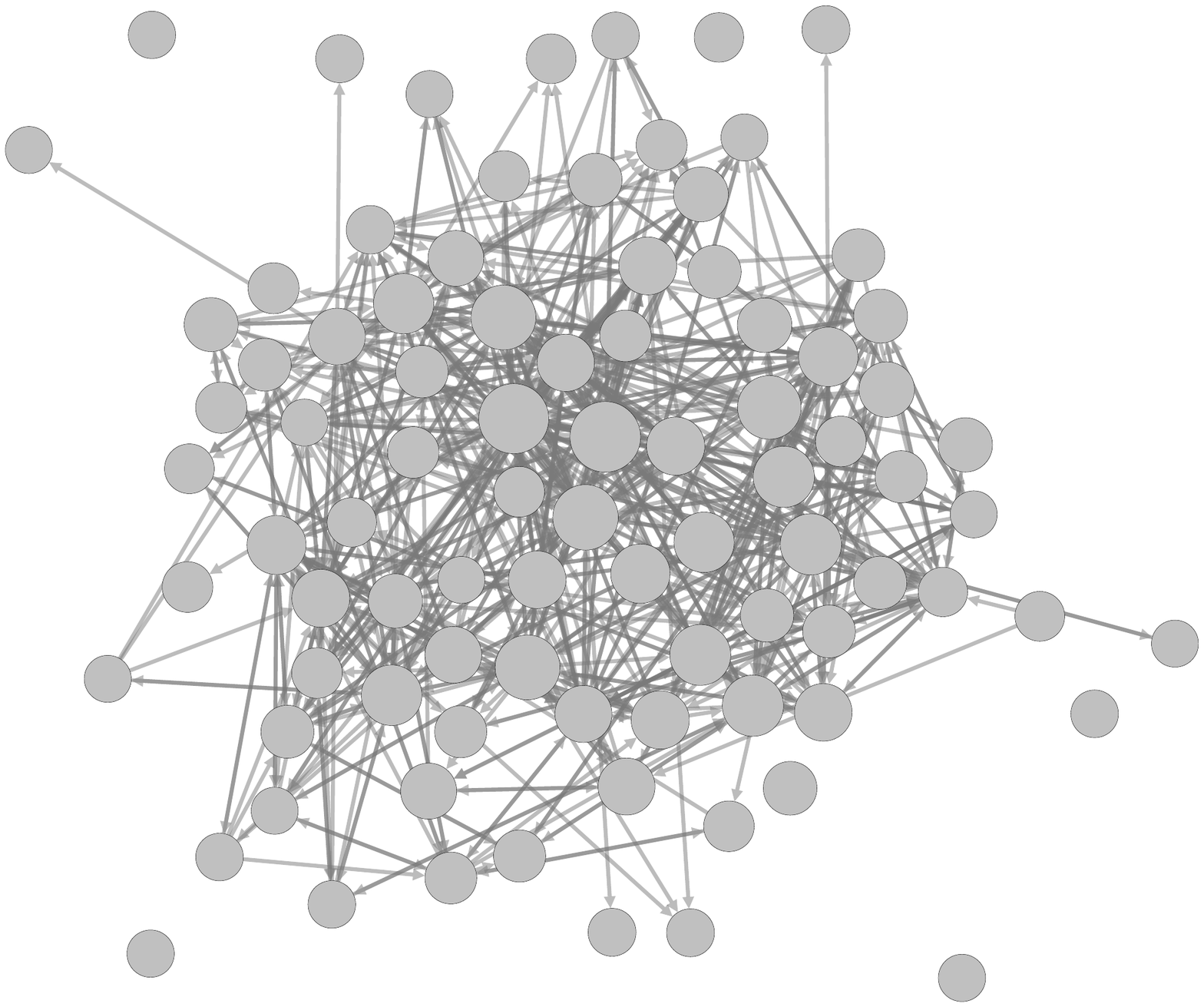}
	\caption{The European research institution email correspondence aggregated to  a binary network divided into   \textsl{Period 1} (day 1-365, left) and \textsl{Period 2} (day 1-365, right). The node size relates to the sum of the out- and in-degree.}
	\label{fig:networks2}
\end{figure}

 The data contains $n=89$ persons and is recorded over $802$ days. For this paper, we select the first two years and split them again into two years, labeled \textsl{Period 1} and \textsl{Period 2}. Within the first period,  $8\,068$ events are recorded and $4\,031$ in the second period. We only regard one-to-one email correspondences, therefore we exclude all group mails from the analysis. In the right column of Table \ref{tbl:descriptive} the descriptive measures for the two aggregated networks are given and in Figure  \ref{fig:networks2} they are visualized. All descriptive statistics are higher in the email exchange network as compared to the arms trade network. In comparison to the arms trade network, the aggregated network is more dense with more than 10\% of all possible ties being realized. In both years the transitivity measure is relatively higher in both time periods. The high share of reciprocated ties is intuitive given that the network represents email exchange between institution members that may collaborate. No covariates are available for this network. See Annex \ref{sec:annex} for the visualization of the degree distributions of both applications. 

\section{Dynamic Exponential Random Graph Models}\label{sec:discrete}
\subsection{Temporal Exponential Random Graph Model}
The Exponential Random Graph Model (ERGM) is among the most popular models for the analysis of static network data. \citet{holland1981} introduced the model class, which was subsequently extended with respect to fitting algorithms and network statistics (see \citealp{lusher2012}, \citealp{robins2007}). Spurred by the popularity of ERGMs, dynamic extensions of this model class emerged, pioneered by \citet{robins2001} who developed time-discrete models for temporally evolving social networks. Before we start with a description of the model, we want to highlight that the TERGM as well as the STERGM are most appropriate for \textsl{equidistant} time points. That is, we observe the networks $Y_t$ at discrete and equidistant time points $t=1,...,T$. Only in this setting, the parameters allow for a meaningful interpretation. See \citet{block2018} for a deeper discussion.

\citet{hanneke2010} is the main reference for the TERGM, a model class that utilizes the Markov structure and, thereby, assumes that the transition of a network from time point $t-1$ to time point $t$ can be explained by exogenous covariates as well as structural components of the present and preceding networks. Using a first-order Markov dependence structure and conditioning on the first network, the resulting dependence structure of the model can be factorized into 
\begin{equation}
\mathbb{P}_\theta(Y_T,...,Y_2|Y_1,x_1,...,x_T)=\mathbb{P}_\theta(Y_T|Y_{T-1},x_T)\cdots \mathbb{P}_\theta(Y_3|Y_2,x_3)\mathbb{P}_\theta(Y_2|Y_1,x_2).
\label{eq:markov}
\end{equation}
In the formulation above, it is assumed that the joint distribution can be decomposed into yearly transitions from $Y_{t-1}$ to $Y_t$. Further,
 it is assumed that the same parameter vector $\theta$ governs all transitions. Often, this is an unrealistic assumption for networks observed at many time-points because the generative process may change other time. Therefore, it can be useful to allow for different parameter vectors for each transition probability (i.e.\ $\theta_T$, $\theta_{T-1},...$). In such a setting, the parameters for each transition can either be estimated sequentially (e.g. \citealp{thurner2018}) or by using smooth time-varying effects (e.g. \citealp{lebacher2019}).

Given the dependence structure (\ref{eq:markov}), the TERGM assumes that the transition from $Y_{t-1}$ to $Y_{t}$ is generated according to an exponential random graph distribution with the parameter $\theta$:
\begin{equation}
\label{eq:ergm}
\mathbb{P}_\theta(Y_t=y_t|Y_{t-1}=y_{t-1},x_t)=\frac{\exp\{\theta^Ts(y_t,y_{t-1},x_t)\}}{\kappa(\theta,y_{t-1},x_t)}.
\end{equation}
Generally, $s(y_t,y_{t-1},x_t)$ specifies a $p$-dimensional function of sufficient network statistics which may depend on the present and previous network as well as on covariates. These network statistics can include static components, designed for cross-sectional dependence structures (see \citealp{morris2008} for more examples). However, the statistics $s(y_t,y_{t-1},x_t)$ explicitly allow temporal interactions, e.g.\ delayed reciprocity
\begin{equation}
\label{eq:recip}
s_{delrecip}(y_t,y_{t-1})\propto \sum_{i\neq j} y_{ji,t}y_{ij,t-1}.
\end{equation} This statistic governs the tendency whether a tie $(i,j)$ in $t-1$ will be reciprocated in $t$. Another important temporal statistic is stability
\begin{equation}
\label{eq:stability}
s_{stability}(y_t,y_{t-1})\propto \sum_{i\neq j}\left( y_{ij,t}y_{ij,t-1}+(1-y_{ij,t})(1-y_{ij,t-1}) \right).
\end{equation}
In this case, the first product in the sum measures whether existing ties in $t-1$ persist in $t$ and the second term is one if non-existent ties in $t-1$ remain non-existent in $t$. The proportionality sign is used since in many cases the network statistics are scaled into a specific interval (e.g.\ $[0,n]$ or $[0,1]$). Such a standardization is especially sensible for networks where the actor set changes with time. Additionally, exogenous covariates can be included, e.g.,  time-varying covariates $x_{ij,t}$
\begin{equation}
\label{eq:exogenous}
s_{dyadic}(y_t,x_{t})= \sum_{i\neq j} y_{ij,t} x_{ij,t}.
\end{equation} 
There exists  an abundance of possibilities for defining interactions between ties in $t-1$ and $t$. From this discussion and equation (\ref{eq:ergm}) it also becomes evident, that in a situation where the interest lies in the transition between two periods, a TERGM can be modeled simply as an ERGM, including lagged network statistics. This can be done for example by incorporating $y_{ij,t-1}$ as explanatory variable in (\ref{eq:exogenous}) which is mathematically equivalent to the stability statistic \eqref{eq:stability}. In the application we call this statistic \textsl{repetition} \citep{block2018}.


Concerning the estimation of the model, maximum likelihood estimation appears to be a natural candidate due to the simple exponential family form  (\ref{eq:ergm}). However, the normalization constant in the denominator of model \eqref{eq:ergm} often poses an inhibiting obstacle when estimating (T)ERGMs. This can be seen by inspecting the normalization constant $\kappa(\theta,y_{t-1},x_t)=\sum_{\tilde{y}\in \mathcal{Y}}\exp\{\theta^Ts(\tilde{y}_t,y_{t-1},x_t)\}$, that requires  summation over \textsl{all possible} networks $\tilde{y}\in \mathcal{Y}$. This task is virtually infeasible, except for very small networks. Therefore, Markov Chain Monte Carlo (MCMC) methods have been proposed in order to approximate the logarithmic likelihood function (see \citet{geyer1992} for Monte Carlo maximum likelihood and \citet{hummel2012} for its adaption to ERGMs). The article by \citet{caimo2011bayesian} provides an alternative algorithm that uses  MCMC-based inference  in a Bayesian model framework.
Another approach is to employ maximum pseudolikelihood estimation (MPLE, \citealp{strauss1990pseudolikelihood}) that can be viewed as a local alternative to the likelihood \citep{vanduijn2009} but   is often regarded as unreliable and poorly understood in the literature (\citealp{Hunter2008a}, \citealp{handcock2003assessing}). However, the MPLE is asymptotically consistent (\citealp{desmarais2012statistical}) and the often suspect standard errors can be corrected via bootstrap (\citealp{leifeld2017}).
 A notable special case arises if the network statistics are restricted such that they decompose to
\begin{equation}
\label{eq:simple}
s(y_t,y_{t-1},x_t)=\sum_{i\neq j}y_{ij,t}\tilde{s}_{ij}(y_{t-1},x_t),
\end{equation}
with $\tilde{s}_{ij}$ being a function that is evaluated only at the lagged network $y_{t-1}$ and covariates $x_t$ for tie $(i,j)$. With this restriction, we impose that the ties in $t$ are independent, conditional on the network structures in $t-1$. This greatly simplifies the estimation procedure and allows to fit the model as a logistic regression model (see for example \citealp{almquist2014}) without the issues related to the MPLE.

A problem, that is very often encountered when fitting (T)ERGMs with endogenous network statistics is called \textsl{degeneracy} (\citealp{schweinberger2011instability}) and occurs if most of the probability mass is attributed to network realizations that provide either full or empty networks. One way to circumvent this problems is  the inclusion of modified statistics, called \textsl{geometrically weighted statistics} (\citealp{snijders2006}). Using the definitions of \citet{hunter2007curved}, the \textsl{geometrically weighted out-degree distribution} (\textsl{GWOD}) controls for the out-degree distribution with one statistic, via 
\begin{equation}
\label{eq:gwo}
s_{GWOD}(y_t)=\exp\{\alpha_O\} \sum_{k=1}^{n-1}\big(1-(1-\exp\{-\alpha_O\})^k\big)O_k(y_t),
\end{equation}
with $O_k(y_t)$ being the number of nodes with out-degree  $k$ in $t$ and $\alpha_O$ as the weighting parameter. Correspondingly, the in-degree distribution is captured by the \textsl{geometrically weighted in-degree distribution} (\textsl{GWID}) statistic by exchanging $O_k(y_t)$ with $I_k(y_t)$, which counts the number of nodes with in-degree $k$, and $\alpha_O$ with  $\alpha_I$. While on the one hand, the weighting often effectively counteracts the problem of degeneracy, the statistics become more complicated to interpret.
Negative values of the associated parameter typically indicate a centralized network structure. 

Regarding statistics capturing clustering, the most common geometrically weighted triangular structure is called \textsl{geometrically weighted edge-wise shared partners} (GWESP) and builds on the number of two-paths that indirectly connect two nodes $i$ and $j$ given the presence of an edge $(i,j)$:
\begin{equation}
\label{eq:gwesp}
s_{GWESP}(y_t)=\exp\{\alpha_S\} \sum_{k=1}^{n-2}\big(1-(1-\exp\{-\alpha_S\})^k\big)S_k(y_t),
\end{equation}
where $\alpha_S$ is a weighting parameter. The number of edges with $k$ shared partners ($S_k(y_t)$) is uniquely defined in undirected networks. If the edges are directed it must be decided, which combination should form a triangle, see \citet{lusher2012} for a discussion.  As a default, the number of directed two-paths is chosen \citep{goodreau2009}. Generally, a positive coefficient for GWESP indicates that triadic closure increases the probability of edge occurrence and globally a positive value for the associated parameter means more triadic closure as compared to a regime with a negative value \citep{morris2008}.

\subsection{Separable Temporal Exponential Random Graph Model}    \label{sec:stergm}

An useful improvement of the TERGM (\ref{eq:ergm}) is the STERGM proposed by \citet{krivitsky2014}. This model can be motivated by the fact that the stability term leads to an ambiguous interpretation of its corresponding parameter. Given that we include (\ref{eq:stability}) in a TERGM and obtain a positive coefficient after fitting the model it is not clear whether the network can be regarded as "stable" because existing ties are not dissolved (i.e. $y_{ij,t}=y_{ij,t-1}=1$) or because no new ties are formed (i.e. $y_{ij,t}=y_{ij,t-1}=0$). To disentangle this, the authors propose a model that allows for the separation of formation and dissolution.

\begin{figure}[t!]
    \centering
    \includegraphics[trim={0cm 0cm 0cm 0cm},clip,width=0.9\textwidth]{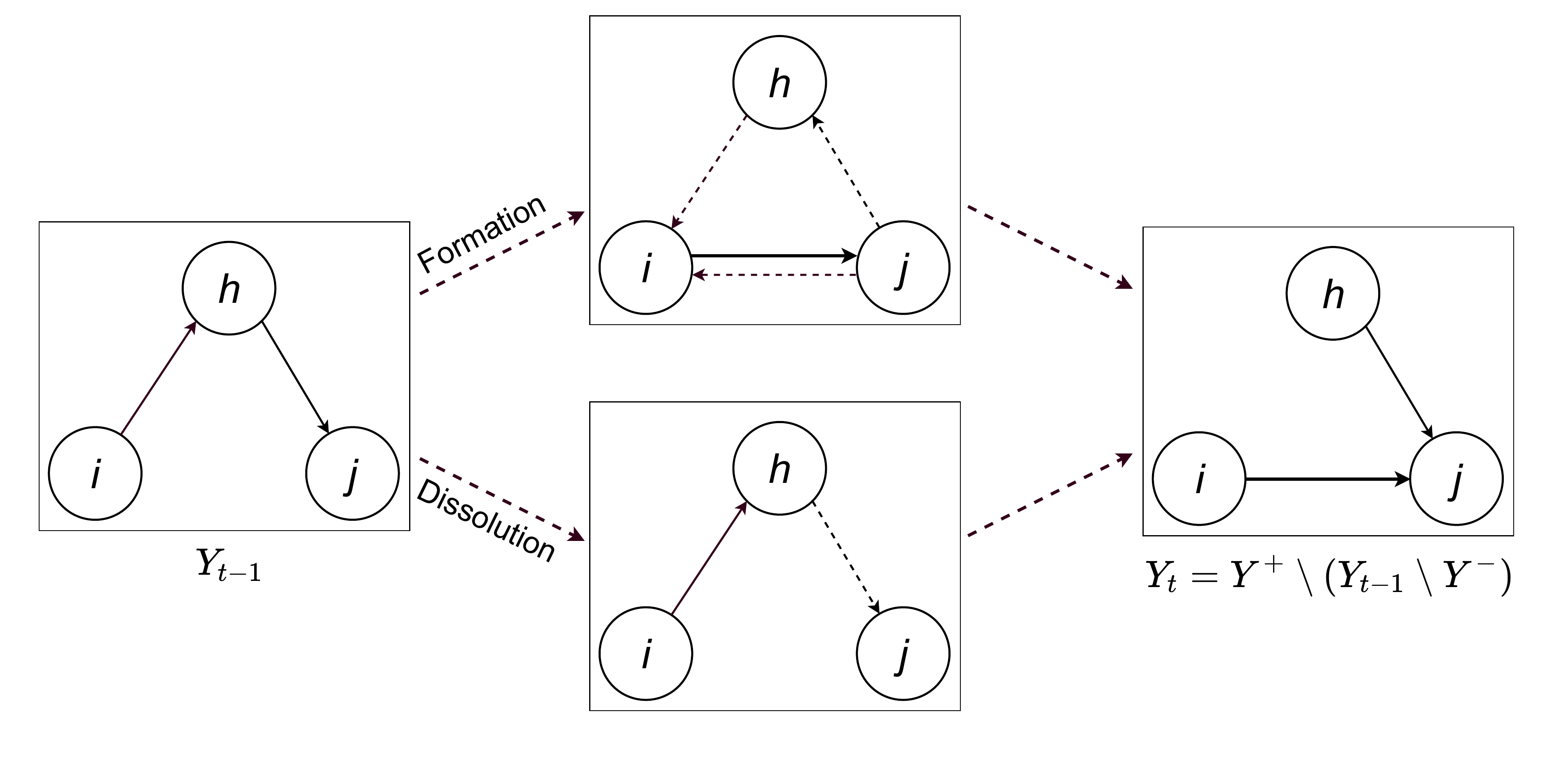}
    \caption{Conceptual representation, illustrating formation and dissolution in the STERGM.}
    \label{fig:stergm}
\end{figure}

\citet{krivitsky2014} define the \textsl{formation network} as $Y^+=Y_t \cup Y_{t-1}$, being the network that consists of the initial network $Y_{t-1}$ together with all ties that are newly added in $t$. The \textsl{dissolution network} is given by $Y^-=Y_t \cap Y_{t-1}$ and contains exclusively ties that are present in $t$ and $t-1$. Given the network in $t-1$ together with the formation and the dissolution network we can then uniquely reconstruct the network in $t$, since $Y_t=Y^+\backslash(Y_{t-1}\backslash Y^-)=Y^-\cup(Y^+\backslash Y_{t-1})$. Define $\theta=(\theta^+,\theta^-)$ as the joint parameter vector that contains the parameters of the formation and the dissolution model. Building on that, \citet{krivitsky2014} define their model to be separable in the sense that the parameter space of $\theta$ is the product of the parameter spaces of $\theta^+$ and $\theta^-$ together with conditional independence of formation and dissolution given the network in $t-1$:
\begin{align}
\label{eq:separate}
&\mathbb{P}_{\theta}(Y_t=y_t|Y_{t-1}=y_{t-1},x_t)= \nonumber \\ & \hspace{0.1cm}\underbrace{\mathbb{P}_{\theta^+}(Y^+=y^+|Y_{t-1}=y_{t-1},x_t)}_{\text{Formation Model}}\underbrace{\mathbb{P}_{\theta^-}(Y^-=y^-|Y_{t-1}=y_{t-1},x_t)}_{\text{Dissolution Model}}.
\end{align}
The structure of the model is visualized in Figure \ref{fig:stergm}. On the left hand side the state of the network $Y_{t-1}$ is given, consisting of two ties $(i,h)$ and $(h,j)$. In the top network all ties that could possibly be formed are shown in dashed and the actual formation in this example $(i,j)$ is shown in solid. On the bottom, the two ties that could possibly be dissolved are shown and in this example ($h,j$) persists while $(i,j)$ is dissolved. On the right hand side of Figure \ref{fig:stergm} the resulting network at time point $t$ is displayed. Given this structure and the  separability assumption (\ref{eq:separate}), it is assumed that the formation model is given by
\begin{equation}
\label{eq:stergm}
\mathbb{P}_{\theta^+}(Y^+=y^+|Y_{t-1}=y_{t-1},x_t)=\frac{\exp\{(\theta^+)^Ts(y^+,y_{t-1},x_t)\}}{\kappa(\theta^+,y_{t-1},x_t)},
\end{equation}
with  $\kappa(\theta^+,y_{t-1},x_t)$ being the normalization constant. Accordingly, the dissolution model can be defined. Inserting the separable models in (\ref{eq:separate}) makes it apparent, that the STERGM is a subclass of the TERGM since 
\begin{align*}
\label{eq:separate}
\mathbb{P}_{\theta}(Y_t=y_t|Y_{t-1}=y_{t-1},x_t)&=  \frac{\exp\{(\theta^+)^Ts(y^+,y_{t-1},x_t)\}}{\kappa(\theta^+,y_{t-1},x_t)} \frac{\exp\{(\theta^-)^Ts(y^-,y_{t-1},x_t)\}}{\kappa(\theta^-,y_{t-1},x_t)} \nonumber\\
&=  \frac{\exp\{\theta^Ts(y_t,y_{t-1},x_t)\}}{\kappa(\theta,y_{t-1},x_t)} 
\end{align*}  
with $\theta=(\theta^+,\theta^-)^T$, $s(y_t,y_{t-1},x_t)=(s(y^+,y_{t-1},x_t),s(y^-,y_{t-1},x_t))^T$ and the normalization constant set accordingly.

For practical reasons it is important to understand that the term \textsl{dissolution} model is somewhat misleading since a positive coefficient in the dissolution model implies that nodes (or dyads) with high values for this statistic are \textsl{less likely} to dissolve. This is also the standard implementation in software packages, but can simply be changed by switching the signs of the parameters in the dissolution model.

The network statistics are used similarly as in a cross-sectional ERGM. In \citet{krivitsky2014} they are called \textsl{implicitly dynamic} because they are evaluated either at the formation network $y^+$ or the dissolution network $y^-$, which are both formed from $y_{t-1}$ and $y_t$. For example, the number of edges is separately computed now for the formation and the dissolution network, giving either the number of edges that newly formed or the number of edges that persisted. 
For example, reciprocity in the formation network is defined as 
\begin{equation}
\label{eq:recip+}
s_{recip}(y^+,y_{t-1})=s_{recip}(y^+)\propto \sum_{i\neq j} y_{ji}^+y_{ij}^+
\end{equation}
and in case of the dissolution model, $y^+$ is simply exchanged with $y^-$.
Similarly, and edge covariates or the geometrically weighted statistics shown in equations (\ref{eq:exogenous}),  (\ref{eq:gwo}) and  (\ref{eq:gwesp}) are now functions of $y^+$ or $y^-$ not $y_t$. 

\subsection{Model Assessment}
\label{sec:goodness-pf-fit-discrete}
In analogy to binary regressions models, the (S)TERGM can be evaluated in terms of their receiver-operator-\newline characteristic (ROC) curve or precision-recall (PR), where the latter puts more emphasis on finding true positives (e.g.\ \citealp{grau2015}). A comparison between different models is possible using, for example, the Akaike Information Criterion (AIC, \citealp{claeskens2008}). Here we want to highlight, that the AIC fundamentally builds on the log likelihood, which in most realistic applications is only available as an approximation, see \citet{Hunter2008a} for further discussion.  

  However, in statistical network analysis it is often argued that suitable network models should not exclusively provide good predictions for individual edges, but also be able to represent topologies of the observed network. The dominant approach to asses the goodness-of-fit of (S)TERGMs is based on sampling networks from their distribution under the estimated parameters and then comparing network characteristics of these sampled networks with the same ones from the observed network (\citealp{Hunter2008a}). For this approach, it is recommendable to utilize network characteristics that are not used for specifying the model. For instance, models that include the GWOD statistic \eqref{eq:gwo} may not be compared to the simulated values of it but against the out-degree distribution.

\citet{hanneke2010} point out that for networks with more than one transition from $t-1$ to $t$ available it is possible to employ a "cross-validation-type" assessment of the fit. The parameters can be fit repetitively to all observed transitions except one hold-out transition. It is then checked, how well the network statistics from the hold-out transition period are represented by the ones sampled from the coefficients obtained from all other transitions. 

\section{Relational Event Model}\label{sec:process}

\subsection{Time-Continuous Event Processes}

The second type of dynamic network models results by comprehending network changes as a continuously evolving process (see \citealp{girardin2018} as a basic reference for stochastic processes). The idea was originally introduced by \citet{holland1977}. According to their view, changes in the network are not occurring at discrete time points but as a continuously evolving process, where only one tie can be toggled at a time. This framework was extended by \citet{butts2008} to model behavior, which is understood as a \textsl{directed event} at a specific time, that potentially depends on the past. Correspondingly, the observations in this section are behaviors which are given as triplets $\omega = (i,j,\tilde{t})$ and encode the sender $i$, receiver $j$, and exact time point $\tilde{t}$. This fine-grained temporal information is often called \textsl{time-stamped} or \textsl{time-continuous}, we adopt the latter name. Furthermore, we only regard dyadic events in this article, i.e., a behavior only includes one sender and receiver. 

The concept of behavior, hereinafter called event, generalizes the classical concept of binary relationships based on graph theory as promoted by \citet{wasserman_faust_1994}. This event framework does not intrinsically assume that ties are enduring over a specific time frame \citep{butts2009,butts2017}. For example in an email exchange network, sending one email at a specific time point is merely a brief event, which does not convey the same information as a durable relationship. Therefore, the time-stamped information cannot adequately be represented in a binary adjacency matrix without having to aggregate the relational data at the cost of information loss \citep{stadtfeld2012}. Nevertheless, a friendship between actor $i$ and $j$ at a given time point can still be viewed as an event that has an one-to-one analogy to a tie in the classical framework.

The overall aim of Relational Event Models (REM, \citealp{butts2008}) is to understand the dynamic structure of events conditional on the history of events \citep{lerner2013}. This dynamic structure, in turn, controls how past interactions shape the propensity of future events. To make this model feasible, we leverage results from the field of time-to-event analysis, or survival analysis respectively (see, e.g., \citealp{kalbfleisch2002} for an overview). 
The central concept of the REM can be motivated by the introduction of a multivariate time-continuous Poisson counting process  
\begin{equation}
\label{eq:count}
N(\tilde{t}) = (N_{ij}(\tilde{t}) \mid i,j \in \lbrace 1, \ldots, n \rbrace),
\end{equation}where $N_{ij}(\tilde{t})$ counts how often actors $i$ and $j$ interacted in $[0,\tilde{t})$.  Note that we indicate continuous time $\tilde{t}$ with a tilde to distinguish from the discrete time setting with $t = 1, 2, \ldots, T$ assumed in the previous section. Process \eqref{eq:count} is characterized by an intensity function $\lambda_{ij}(\tilde{t})$ for $i \neq j$, which is defined as:
\begin{align*}
\lambda_{ij}(\tilde{t})= \underset{dt \downarrow 0}{\text{lim}} \dfrac{\mathbb{P}(N_{ij}(\tilde{t}+dt) = N_{ij}(\tilde{t}) +1)}{dt}.
\end{align*}
This is the instantaneous probability of observing a jump of size "1" in $N_{ij}(\tilde{t})$, which indicates observing the event $(i,j, \tilde{t})$. Since we assume that there are no self-loops $\lambda_{ii}(\tilde{t})\equiv 0~  \forall~  i = 1, \ldots, n$ holds. 


\subsection{Time-Continuous Observations}
\label{sec:rem}

\citet{butts2008} introduced the REM to analyze the intensity $\lambda_{ij}(\tilde{t})$ of process \eqref{eq:count} when time-continuous data on the events are available. He assumed that the intensity is constant over time but depends on time-varying relational information of past events and exogenous covariates. \citet{vu2011} extended the model by  postulating a semi-parametric intensity similar to \citet{cox1972}: 
\begin{align}
\lambda_{ij}(\tilde{t} \mid N(\tilde{t}), x(\tilde{t}), \theta) = \lambda_0(\tilde{t})\text{exp}\big\lbrace \theta^T s_{ij}\big( N(\tilde{t}), x(\tilde{t})\big) \big\rbrace, 
\label{eq:rem}
\end{align}  
where $\lambda_0(\tilde{t})$ is an arbitrary baseline intensity, $\theta \in \mathbb{R}^p$ the parameter vector and $s_{ij}\big(N(\tilde{t}), x(\tilde{t})\big)$ a statistic that depends on the (possibly time-continuous) covariate process $x(\tilde{t})$ and the counting process just prior to $\tilde{t}$. 

Generally, similar statistics as already introduced in Section \ref{sec:discrete} can be included in \newline $s_{ij}\big( N(\tilde{t}), x(\tilde{t})\big)$. Solely the differing level of the model needs to be accounted for, since model \eqref{eq:rem} takes a local time-continuous point of view to understand the relational nature of the observed events. This necessitates defining the statistics $s_{ij}\big( N(\tilde{t}), x(\tilde{t})\big)$ from the position of specific ties, in contrast to the globally defined statistics $s(y_t,y_{t-1},x_t)$ in \eqref{eq:ergm}. To give an example, the tie-level version of reciprocity for the event $(i,j)$ is defined as
\begin{align*}
s_{ij,reciprocity}\big(N(\tilde{t}), x(\tilde{t})\big) = \mathbb{I}\big( N_{ji}(\tilde{t})> 0 \big),
\end{align*}
where $\mathbb{I}(\cdot)$ is the indicator function. It only regards, whether already having observed the event $(j,i)$ prior to $\tilde{t}$ has an effect on $\lambda_{ij}(\tilde{t} \mid N(\tilde{t}), x(\tilde{t}), \theta)$, in comparison to the network level version \eqref{eq:recip} of delayed reciprocity that counted all reciprocated ties between the networks $y_t$ and $y_{t-1}$. 

Degree statistics can be specified as either sender- or receiver-specific. If we, e.g., want to control for the out-degree of the sender the corresponding tie-oriented statistic is: 
\begin{align*}
s_{ij,SOD}\big(N(\tilde{t}), x(\tilde{t})\big) = \sum_{h = 1}^n \mathbb{I}\big( N_{ih}(\tilde{t}) > 0 \big).
\end{align*} The in-degree of the receiver can be formulated accordingly.

Clustering in event sequences may be captured by different types of nested two-path configurations. For instance, the tie-oriented version of directed two-paths, henceforth called \textsl{transitivity}, is given by: 
\begin{align*}
s_{ij,TRA}\big(N(\tilde{t}), x(\tilde{t})\big) = \sum_{h= 1}^n \mathbb{I}\big( N_{ih}(\tilde{t})> 0 \big) \mathbb{I}\big( N_{hj}(\tilde{t}) > 0 \big).
\end{align*}
The inclusion of monadic and dyadic exogenous covariates becomes straightforward by setting $s_{ij,dyadic}\big(N(\tilde{t}), x(\tilde{t})\big)$ equal to the covariate values of interest. Since the effect of a past event at time $\delta$, say, on a present event at time $\tilde{t}$ may vary according to the elapsed time $\tilde{t} - \delta$, \citet{stadtfeld2017_4} introduced windowed effects, which only regard events that occurred in a pre-specified time window, e.g. a year. We will come back to this point in the next section. 

If time-continuous observations are available each dimension of the observed counting process is conditional on the past independent. This, in turn, enables the construction of a likelihood, which can subsequently be maximized. Assuming that $\Omega$ is the set of all observed events and $\mathcal{T}$ the interval of observation, the likelihood can be written as: 
\begin{align}
\mathcal{L}(\theta)  = \prod_{(i,j,\tilde{t}) \in \Omega} \lambda_{i j}(\tilde{t} \mid N(\tilde{t}), x(\tilde{t}), \theta) \text{exp}\Big\lbrace -\underset{\mathcal{T}}{\int} \sum_{k,h = 1}^n  \lambda_{kh}(u \mid N(u), x(u), \theta) d u\Big\rbrace.\label{eq:c_lh}
\end{align} This likelihood is straightforward to maximize in the case of a parametric baseline intensity $\lambda_0(t)$, for example \citet{butts2008} assumes $\lambda_0(t) = \gamma_0$.
Alternatively, \citet{butts2008} analyzed events with ordinal temporal information. In this setting, the likelihood is equal to the partial likelihood introduced by \citet{cox1972} for estimating parameters of semi-parametric intensities as in \eqref{eq:rem}. Letting $\mathit{U}_t$ denote the set of all possible events that could have occurred at time point $t$ but did not, the partial likelihood for continuous event data is defined as: 
\begin{align}
\mathcal{PL}_{cont}(\theta)  = \prod_{(i,j,\tilde{t})  \in \Omega} \frac{\lambda_{i j}(\tilde{t}\mid N(\tilde{t}), x(\tilde{t}), \theta)}{\sum_{(k,h) \in \mathit{U}_{\tilde{t}}} \lambda_{kh}(\tilde{t} \mid N(\tilde{t}), x(\tilde{t}), \theta)}.\label{eq:cpartial_lh}
\end{align} Consecutively, $\Lambda_0(t) = \int_0^t \lambda_0(u) du$ can be estimated with a Nelson Aalen estimator (see \citealp{kalbfleisch2002} for further details on the estimation). 

When dealing with large amounts of event data the main obstacle is evaluating the sum over the intensities of all possible ties in \eqref{eq:cpartial_lh} \citep{butts2008}. One exact option is to trade a longer running time for a slimmer memory footprint by means of a coaching data structure. \citet{vu2011} exploit this by saving prior values of the sum and subsequently changing it event-wise by elements of $\mathit{U}_{\tilde{t}}$ whose covariates changed. Alternatively, \citet{vu2015} proposes approximate routines that utilize case-control sampling and stratification for the Cox model \citep{langholz1995}. More precisely, the sum is only calculated over a sampled subset of possible events in addition to stratification. \citet{lerner2019reliability} go one step further and sample events out of $\Omega$ for the calculation of $\mathcal{PL}(\theta)$ in \eqref{eq:cpartial_lh}. 

Extensions of this model building on already well-established methods in social network and time-to-event analysis were numerously proposed. \citet{perry2013} used a stratified Cox model in \eqref{eq:rem}. \citet{stadtfeld2017_1} adopted the Stochastic Actor oriented Model (SAOM) to events. \citet{dubois2010} and \citet{dubois2013} extended the Stochastic Block Model (SBM) for time-stamped relational events. Further, \citet{dubois2013_1} adopted a Bayesian hierarchical model to event data when information is only available in smaller groups.

\subsection{Time-Clustered Observations}
\label{Time-Clustered Observations}

Generally, the approach discussed above requires time-continuous network data, meaning that we observe the precise time points of all events. To give an instance, in the first data example, this means that we need the exact time point $\tilde{t}$ of an arms trade between country $i$ and $j$. Often, such exact time-stamped data are not available and, in fact, trading between states can hardly be stamped with a single time point $\tilde{t}$. Indeed, we often only observe the time-continuous network process at discrete time points $t = 1, \ldots, T$. In such setting, we may assume a Markov structure in that we do not look at the entire history of the process $N(\tilde{t})$  but just condition the intensity \eqref{eq:rem} on the history of events from the previous observation $t-1$ to $\tilde{t}$.  Technically this means that $N(t)$ is adapted to $\tilde{Y}(\tilde{t}) := N(\tilde{t})-N(t-1)$ and $x(\tilde{t})$ for $\tilde{t} \in [t-1,t]$. We then reframe \eqref{eq:rem} as: 
\begin{align}
\lambda_{ij}(\tilde{t} \mid \tilde{Y}(\tilde{t}), x(\tilde{t}), \theta) = \lambda_0(\tilde{t})\text{exp}\big\lbrace \theta^T s_{ij}\big( \tilde{Y}(\tilde{t}), x(\tilde{t})\big) \big\rbrace. \label{eq:intensity_clust}
\end{align}
In other words, we assume that the intensity of events between $t-1$ and $t$ does not depend on states of the multivariate counting process \eqref{eq:count} prior to $t-1$. For this reason, all endogenous statistics introduced in Section \ref{sec:rem} are now evaluated on $\tilde{Y}(\tilde{t})$ instead of $N(\tilde{t})$. This is a reasonable assumption, if one is primarily interested in short-term dependencies between the individual counting processes. It enables a meaningful comparison to the models from Section \ref{sec:discrete} that assume an analog discrete Markov property. However, we want to emphasize that this dependence structure is not vital to inferential results. 

If we observe the continuous process at discrete time points it is inevitable that we observe time clustered observations, meaning that two or more events happen at the same time point. Under the term \textsl{tied} observations this phenomenon is well known in time-to-event analysis and treated with several approximations. One option is the \textsl{so-called} Breslow approximation (see \citealp{peto1972,breslow1974}). Let therefore
\begin{equation*}
\mathit{O}_t = \lbrace (i,j) \mid N_{ij}(t) - N_{ij} (t-1) >0 \rbrace
\end{equation*}
where element $(i,j)$ is replicated $N_{ij}(t)- N_{ij}(t-1)$ times in $\mathit{O}_t$, that is if an event between $i$ and $j$ occurred multiple times in the interval from $t-1$ to $t$ then $(i,j)$ appears respective times in $\mathit{O}_t$. Given that we have not observed the exact time point of an event we also get no information on the baseline intensity $\lambda_0(\tilde{t})$ in \eqref{eq:rem} for $\tilde{t} \in [t-1,t]$ so that the model simplifies to a discrete choice model structure (see, e.g., \citealp{train2009}) which resembles the partial likelihood \eqref{eq:cpartial_lh} and is defined as: 
\begin{align}
\mathcal{PL}_{clust}(\theta) = \prod_{t = 1}^T \frac{\prod_{(i,j) \in \mathit{O}_t} \text{exp} \lbrace \theta^T s_{ij} \big( \tilde{Y}(t), x(t)\big) \rbrace}{\big(\sum_{(k,h) \in \mathit{U_t} } \text{exp} \lbrace \theta^T s_{kh} \big( \tilde{Y}(t), x(t)\big) \rbrace \big)^{n_t}}, \label{eq:partial_lh}
\end{align}
where $n_t = |\mathit{O}_t|$. Alternatively, one can replace the denominator in \eqref{eq:partial_lh} by considering all possible orders of the unobserved events in $\mathit{O}_t$ giving the average likelihood as introduced by \citet{kalbfleisch2002}. Since this can be a combinatorial and hence numerical challenge, random sampling of time point orders among the time-clustered observations can be used with subsequent averaging, which we call Kalbfleisch-Prentice approximation (see \citealp{kalbfleisch2002}). Further techniques to deal with unknown time ordering are augmenting the clustered events into possible paths of ordered events and adapting the maximum likelihood estimation proposed for the SAOM by \citet{snijders2010} or using random sampling of the ordering. This can be legitimized in cases where we may assume independence among events happening in one year, since the events take a long time to materialize \citep{snijders2017_2}. 


\subsection{Model Assessment}
\label{Goodness-of-Fit-REM}

In comparison to the assessment for models operating in discrete time, widely accepted methods dealing with relational event data are scarce. 
The proposals either stem from time-to-event analysis or regard link prediction, which is the task of predicting the most likely next event given the history of past events \citep{nowell2007}. One example of the former option is the usage of \textsl{Schönfeld residuals} by \citet{vu2017} to check the assumption of proportional intensities, which is central to semi-parametric models as the one proposed by \citet{cox1972}. For the latter approach, we need to define a predictive measure that quantifies how well the next event is predicted. \citet{vu2011} proposed the recall measure that estimates the percentage of test events which are in the list of $K$ most likely next events according to a given model. Evaluating this percentage for different values of $K$ permits a visualization of the predictive capabilities of the model. The strength of the predicted intensity allows the ordering of events according to the probability of being observed next. If we model the propensity of time-clustered events that represent binary adjacency matrices one can alternatively adopt the analysis of the ROC and PR curve introduced in Section \ref{sec:goodness-pf-fit-discrete}. 

\section{Application}
\label{Application} 

When it comes to software, there exist essentially three main \texttt{R} packages that are designed for fitting TERGMs and STERGMs. Most important is the extensive \texttt{statnet} library (\citealp{goodreau2008}) that allows for simulation-based fitting of ERGMs. The library contains the package \texttt{tergm} with implemented methods for fitting STERGMs using MCMC approximations of the likelihood. However, currently the package \texttt{tergm} (version 3.5.2) does not allow for fitting STERGMs with time-varying dyadic covariates for more than two time periods jointly. The package \texttt{btergm} (\citealp{leifeld2017}) is designed for fitting TERGMs using either maximum pseudo-likelihood or MCMC maximum likelihood estimation routines. In order to obtain Bayesian Inference in ERGMs, the package \texttt{bergm} by \citet{JSSv061i02} can be used. Besides implementations in \texttt{R}, the stand-alone program \texttt{PNet} \citep{wang2006pnet} allows for simulating, fitting and evaluating (T)ERGMs. In order to ensure comparable estimates we estimate the TERGM, as well as the STERGM, with the \texttt{statnet} library, using MCMC-based likelihood estimation techniques. We use the package \texttt{ergm} and include delayed reciprocity and the repetition of previous ties as dyadic covariates. The STERGM is fitted using the \texttt{tergm} package.

\citet{marcum2015} implemented the R package   $\mathtt{relevent}$ (version 1.0-4) to estimate the REM for time-stamped data. It was followed by the package $\mathtt{goldfish}$ (version 1.2) by \citet{stadtfeld2018} for modeling event data with precise and ordinal temporal information with an actor- and tie-oriented variant of the REM. Furthermore, it is highly customizable in terms of endogenous and exogenous user terms and will be used in the following applications. 
 


We want to remark that the STERGM coefficients are implicitly dynamic, while in the TERGM all network statistics except the lagged network and delayed reciprocity terms are evaluated on the network in $t$. All covariates of the REM are continuously updated and the intensity at time point $\tilde{t}\in [t-1,t]$ only depends on events observed in $[t-1, \tilde{t})$. Like the \textsl{building period} proposed by \citet{vu2011_2}, the events in $t-1$ are only used for building up the covariates and not directly modeled. Due to no compositional changes, we did not scale any statistics. Moreover, we refer to the Supplementary Material for the model assessment.


\subsection{Data Set 1: International Arms Trade}
\label{sec:data1}

\begin{table}[th!]
	\begin{center}\small
		\begin{tabular}{l c c c | c r }
			\hline
			& TERGM &  \multicolumn{2}{c|}{STERGM} & REM & \\ 
			&  & Formation & Dissolution & & \\ 
			\hline \hline 
			
			Repetition                & $3.671^{***}$   & $-$   & $-$ &  $2.661^{***}$ &   \\
			& $(0.132)$       & $-$       & $-$   & $(0.143)$ &    \\
			Edges                          & $-15.632^{***}$ & $-17.186^{***}$ & $-16.987^{***}$& $-$ &  \\
			& $(1.809)$       & $(2.168)$       & $(3.587)$ & $-$ &       \\
			Reciprocity                 & $-0.258$        & $-0.620$        & $-0.058$   & $ -0.109$ &      \\
			& $(0.306)$       & $(0.436)$       & $(0.619)$   & $(0.181)$ &     \\
			In-Degree (GWID) & $-1.823^{***}$  & $-2.106^{***}$  & $-0.412$    &$0.060^{**}$  & In-Degree Receiver     \\
			& $(0.278)$       & $(0.379)$       & $(0.442)$    &  $(0.015)$  &    \\
			Out-Degree (GWOD) & $-3.220^{***}$  & $-4.126^{***}$  & $-0.326$    & $0.010^{**}$ &  Out-Degree Sender   \\
			& $(0.304)$       & $(0.462)$       & $(0.533)$   & $(0.004)$&     \\
			GWESP  & $0.050$         & $0.076$         & $0.150$    & $0.010$ &    Transitivity \\
			& $(0.066)$       & $(0.071)$       & $(0.126)$   & $(0.029)$ &     \\
			Polity Score & $-0.024^{*}$    & $-0.028^{*}$    & $-0.016$  & $-0.016$ &       \\
			 & $(0.010)$       & $(0.014)$       & $(0.017)$   & $(0.009)$ &    \\
			log(GDP) Sender                 & $0.313^{***}$   & $0.394^{***}$   & $0.323^{***}$ & $0.395^{***}$ &   \\
			& $(0.048)$       & $(0.054)$       & $(0.088)$   & $(0.039)$ &     \\
			log(GDP) Receiver                  & $0.165^{***}$   & $0.135^{*}$     & $0.327^{***}$   & $0.192^{***}$ & \\
			& $(0.043)$       & $(0.054)$       & $(0.087)$    & $(0.032)$ &    \\
			
			\hline
			Log Likelihood & $-$949.833        & $-$675.327        & $-$258.425       &  &  \\
			AIC          &  1917.666      & 1366.654        & 532.849   &  &    \\
			$\sum$ AIC             &1917.666         &\multicolumn{2}{c|}{  1899.503}      &   &   \\
			\hline
		\end{tabular}
		\caption{Arms trade network: Comparison of parameters obtained from the  TERGM (first column), STERGM (Formation in the second column, Dissolution in the third column) and REM (fourth column). Standard errors in brackets and stars according to $p$-values smaller than $0.001$ ($^{***}$), $0.05$ ($^{**}$) and $0.1$ ($^{*}$). Decay parameter of the geometrically weighted statistics is set to $\log(2)$ and the Kalbfleisch-Prentice approximation was used with 100 random orderings of the events to find the estimates of the REM.}
		\label{table:coefficients1}
	\end{center}
\end{table}

The results obtained for the arms trading data section are displayed in Table \ref{table:coefficients1}. For a detailed interpretation of effects  focusing on political, social, and economic aspects we refer to the relevant literature (e.g. \citealp{thurner2018}). Here we
want to comment on a few aspects only. 
While we do not have timestamps for the arms trades, the longitudinal networks can still be viewed as time-clustered observations enabling the techniques from Section \ref{Time-Clustered Observations}. 

Both, the TERGM (column 1) and the REM (column 4) identify the repetition of previous ties as a driving force in the dynamic structure of the network. 
Degree-related covariates, which are GWID and GWOD in the (S)TERGM and in- and out-degree in the REM, capture centrality in the network. The coefficients of the GWID and GWOD are negative and have low $p$-values in the TERGM. This stands in contrast to the STERGM, where these effects are only pronounced in the formation model (column 2), while they are insignificant effect in the dissolution model (column 3). Hence, these effects suggest  a centralized pattern in the formation network, which is also captured by the TERGM. In the REM an analogous pattern can be detected, since a higher in-degree of the receiver increases the respective intensity, thus spurs trade relations. Similar interpretations hold for the out-degree of the sender. Overall, countries that have a high out-degree are more likely to send weapons and countries with a high in-degree to receive weapons, which again results in a centralized network structure as indicated by the estimates in the TERGM and STERGM. 


Lastly, consistent effects among the models were also found for the exogenous covariates. Consider, for instance, the coefficient of the logarithmic GDP of the importing country. The TERGM assigns a significantly higher probability to observe in-going ties to countries with a high GDP just like the REM. However, disentangling the model towards formation and dissolution we see strongly significant coefficients in the dissolution model while the effect for the formation model is weakly significant. 

Based on the independence assumption in \eqref{eq:separate} we can sum up the two AIC values and see that the AIC value of the STERGM is smaller than of the TERGM.


\subsection{Data Set 2: European Research Institution Email Correspondence}

\begin{table}[th!]
	\begin{center}\small
		\begin{tabular}{l c c c | c r }
			\hline
			& TERGM &  \multicolumn{2}{c|}{STERGM} & REM & \\ 
			&  & Formation & Dissolution & & \\ 
			\hline \hline 
			
			Repetition              & $1.367^{***}$  & $-$  & $-$   & $2.27^{***}$ & \\
			& $(0.107)$      & $-$      & $-$      & $(0.084)$ &  \\
			Edges                          & $-5.755^{***}$ & $-4.853^{***}$ & $-2.237^{***}$   & $-$ &\\
			& $(0.237)$      & $(0.247)$      & $(0.224)$       & $-$ & \\
			Reciprocity                 & $0.398^{***}$  & $2.498^{***}$  & $2.586^{***}$    & $1.655^{***}$ &\\
			& $(0.112)$      & $(0.157)$      & $(0.226)$    & $(0.075)$ &    \\
			In-degree (GWID) & $1.060^{**}$   & $1.349^{*}$    & $0.709$       & $-0.004$ & In-Degree Receiver     \\
			& $(0.333)$      & $(0.648)$      & $(0.415)$      & $(0.003)$ &  \\
			Out-degree (GWOD) & $0.031$        & $-0.411$       & $-0.369$   & $-0.0001$ & Out-Degree Sender     \\
			& $(0.312)$      & $(0.431)$      & $(0.397)$     & $(0.003)$ &   \\
			GWESP  & $1.560^{***}$  & $0.655^{***}$  & $0.429^{***}$   & $0.070^{***}$&    Transitivity  \\
			& $(0.110)$      & $(0.111)$      & $(0.086)$      & $(0.008)$ &  \\
			
			\hline
			Log Likelihood &    $-$1723.732      &  $-$1000.506      &  $-$505.431        &  & \\
			AIC          & 3459.464       & 2011.012       & 1020.862      &  &   \\
			$\sum$ AIC             &3459.464         &\multicolumn{2}{c|}{  3031.874}      &  &    \\
			\hline
		\end{tabular}
		\caption{Email exchange network: Comparison of parameters obtained from the  TERGM (first column), STERGM (Formation in the second column, Dissolution in the third column) and REM (fourth column). Standard errors in brackets and stars according to $p$-values smaller than $0.001$ ($^{***}$), $0.05$ ($^{**}$) and $0.1$ ($^{*}$). Decay parameter of the geometrically weighted statistics is set to $\log(2)$.}
		\label{table:coefficients2}
	\end{center}
\end{table}

As already indicated by the descriptive statistics in Table \ref{tbl:descriptive}, the email network seems to be driven by three major structural influences: repetition, reciprocity, and transitive clustering. The estimates from Table \ref{table:coefficients2} demonstrate, that all models were able to identify these forces.

According to the REM (column 4), the event network of email traffic in the research institution is not centralized and primarily based on collaboration between coworkers. We can draw those conclusions from insignificant estimates of degree-related statistics and highly significant estimates regarding reciprocity and repetition. In the TERGM (column 1) we find a positive and significant effect of GWID, while no  effect can be found in the STERGM (columns 2 and 3). The estimates of repetition and reciprocity in the REM and TERGM are very pronounced. For instance, the estimates of the REM imply that a reciprocated event is 19.6 times more likely than an event with the same covariates only not being reciprocated. Interestingly, the STERGM detects a lower effect of GWESP in the formation and dissolution than the TERGM. The effect of the delayed reciprocity in the TERGM is less relevant than reciprocity in the formation and dissolution model. This strongly differing effect size results from the mathematical formulation of the statistics given in equations \eqref{eq:recip} and \eqref{eq:recip+}. 

Contrasting the AIC values of the TERGM and STERGM shows that the dynamic structure of the email network is again better explained by the STERGM. In the Supplementary Material we fit the TERGM and STERGM to multiple time points.

\section{Conclusion} \label{sec:discuss}   

\subsection{Further Models}
	\citet{snijders1996} formulated a  two-stage process model operating in a continuous-time framework. The dynamics are considered to evolve according to unobserved micro-steps. At first, a sender out of all eligible actors gets the opportunity to change the state of all his outgoing ties. Consecutively, the actor needs to evaluate the probability of changing the present configuration with each possible receiver, which entails each actors knowledge of the complete graph whenever he has the possibility to toggle one of his ties. Lastly,  the decision is randomly drawn relative to the probabilities of all possible actions. In general, the SAOM is a well-established model for the analysis of social networks, that was successfully applied to a wide array of network data, e.g., in Sociology \citep{agneessens2012,denooy2002}, Political Science \citep{kinne2016,bichler2014}, Economics \citep{castro2014}, and Psychology \citep{jason2014}. Estimation of this model variant is predominantly carried out with the R package $\mathtt{RSiena}$ \citep{Siena}. 
	
	Another notable model that can be regarded as a bridge between the ERGM and continuous-time models is the Longitudinal ERGM (LERGM, \citealp{snijders2013,koskinen2015}). In contrast to the TERGM, the LERGM assumes that the network evolves in micro-steps as a continuous time Markov process with an ERGM being its limiting distribution. Similar to the SAOM, the model builds on randomly assigning the opportunity to change, followed by a function that governs the probability of a tie change.  This model is still tie-oriented, meaning that dyadic ties instead of actors are chosen and then have the option to change the current network.  
	
\subsection{Summary}

In this article, we put emphasis on tie-oriented dynamic network models. Comparisons between these models can be drawn on the level at which each implied generating mechanism works and how time is perceived. The overall aim in the TERGM is to find an adequate distribution of the adjacency matrix $Y_t$ conditioning on information of previous realizations of the network. In the separable extension, the aim remains unchanged, only splitting $Y_t$ into two smaller sub-networks that include all possible ties that were and were not present in $Y_{t-1}$ separately. While the (S)TERGM proceeds in discrete time, the REM tackles modeling the intensity on the tie level in continuous time conditional on past events. Therefore, the TERGM and STERGM take a global and REM a local point-of-view, which results in substantially different interpretations of the estimates. 


Furthermore, we analyzed two data sets that represent two types of network data that are traditionally either modeled by the TERGM and REM. By extending the REM to time-clustered observations and aggregating events to binary adjacency matrices a meaningful comparison between the STERGM, TERGM, and REM is enabled.

		\section*{Acknowledgement}
We thank the anonymous reviewers for their careful reading and constructive comments.
The project was supported by the European Cooperation in Science and Technology [COST Action CA15109 (COSTNET)]. We also gratefully acknowledge funding provided by the German Research Foundation (DFG) for the project  KA 1188/10-1: \textit{International Trade of Arms: A Network Approach}. Furthermore we like to thank the Munich Center for Machine Learning (MCML) for funding.

\bibliographystyle{chicago}	
	\bibliography{literature}
	\appendix

\newpage

\pagenumbering{Roman}

	\section{Annex: Additional Descriptives}
	\label{sec:annex}   
	
\begin{figure}[ht!]\centering
		\includegraphics[trim={0cm 0cm 0cm 0cm},clip,width=0.45\textwidth]{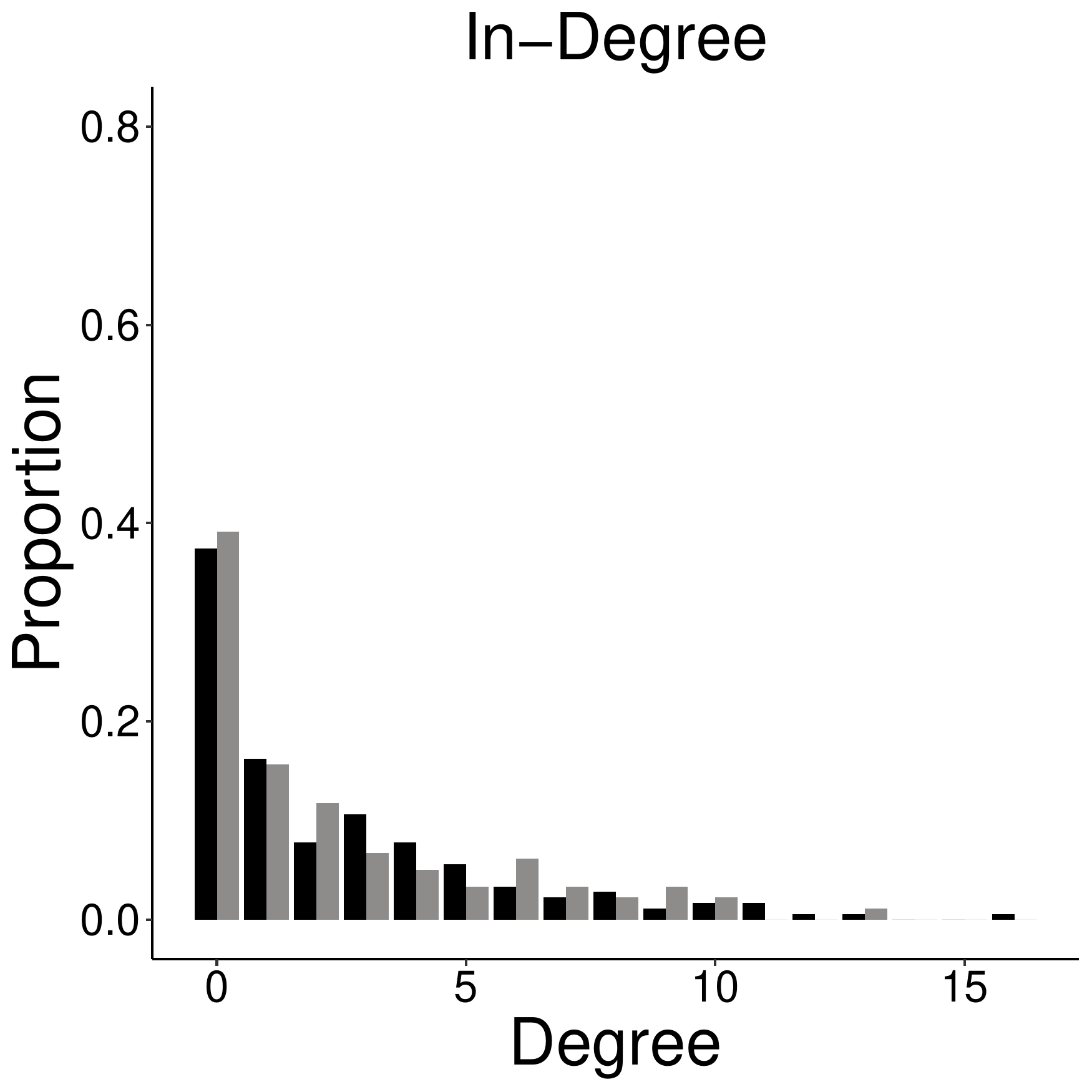}
		\includegraphics[trim={0cm 0cm 0cm 0cm},clip,width=0.45\textwidth]{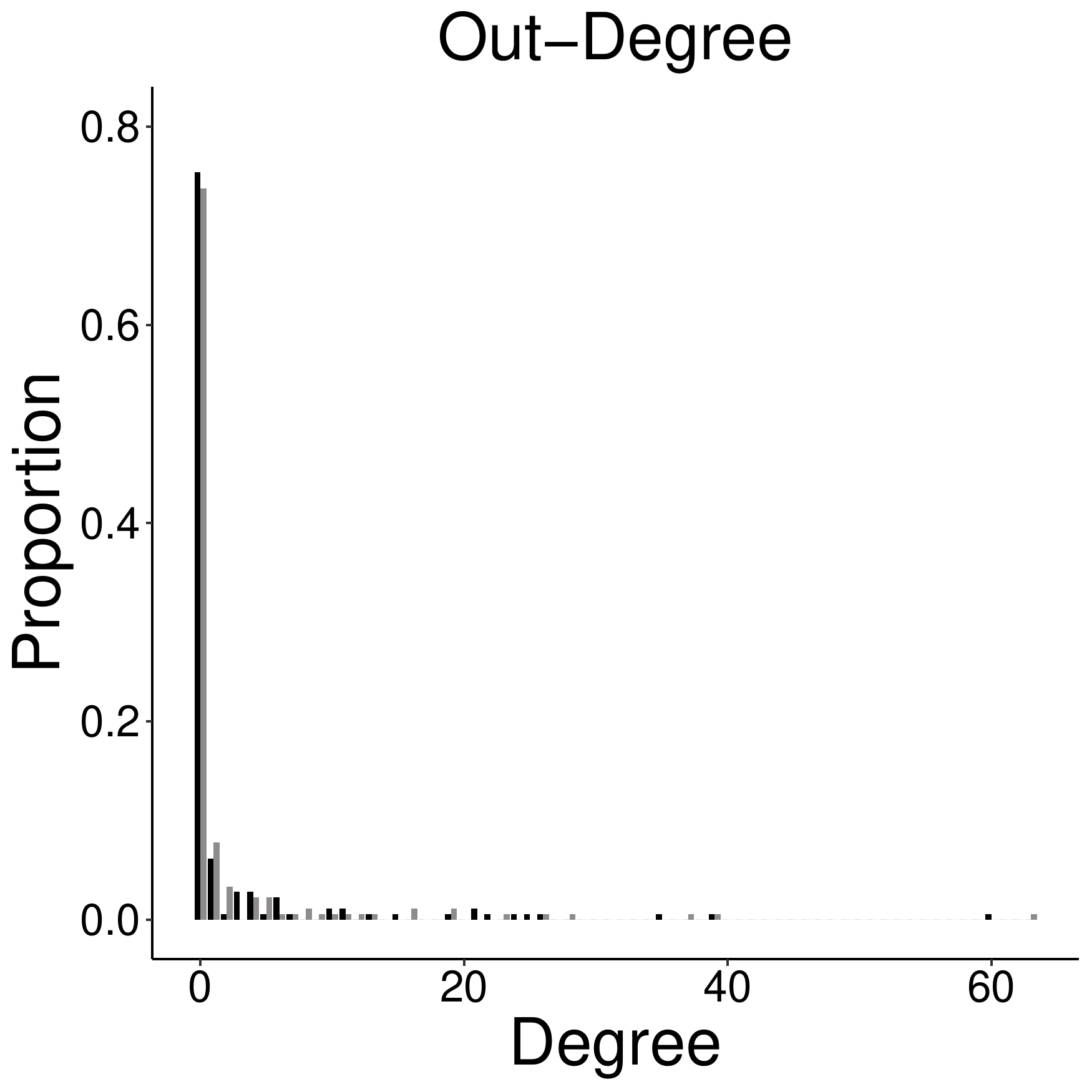}
		\caption{Arms trade network: Barplots indicating the distribution of the in- and out-degrees. Black bars indicate the values of year 2016 and grey bars of 2017.}
		\label{fig:degree}
	\end{figure}

\begin{figure}[ht!]\centering
		\includegraphics[trim={0cm 0cm 0cm 0cm},clip,width=0.45\textwidth]{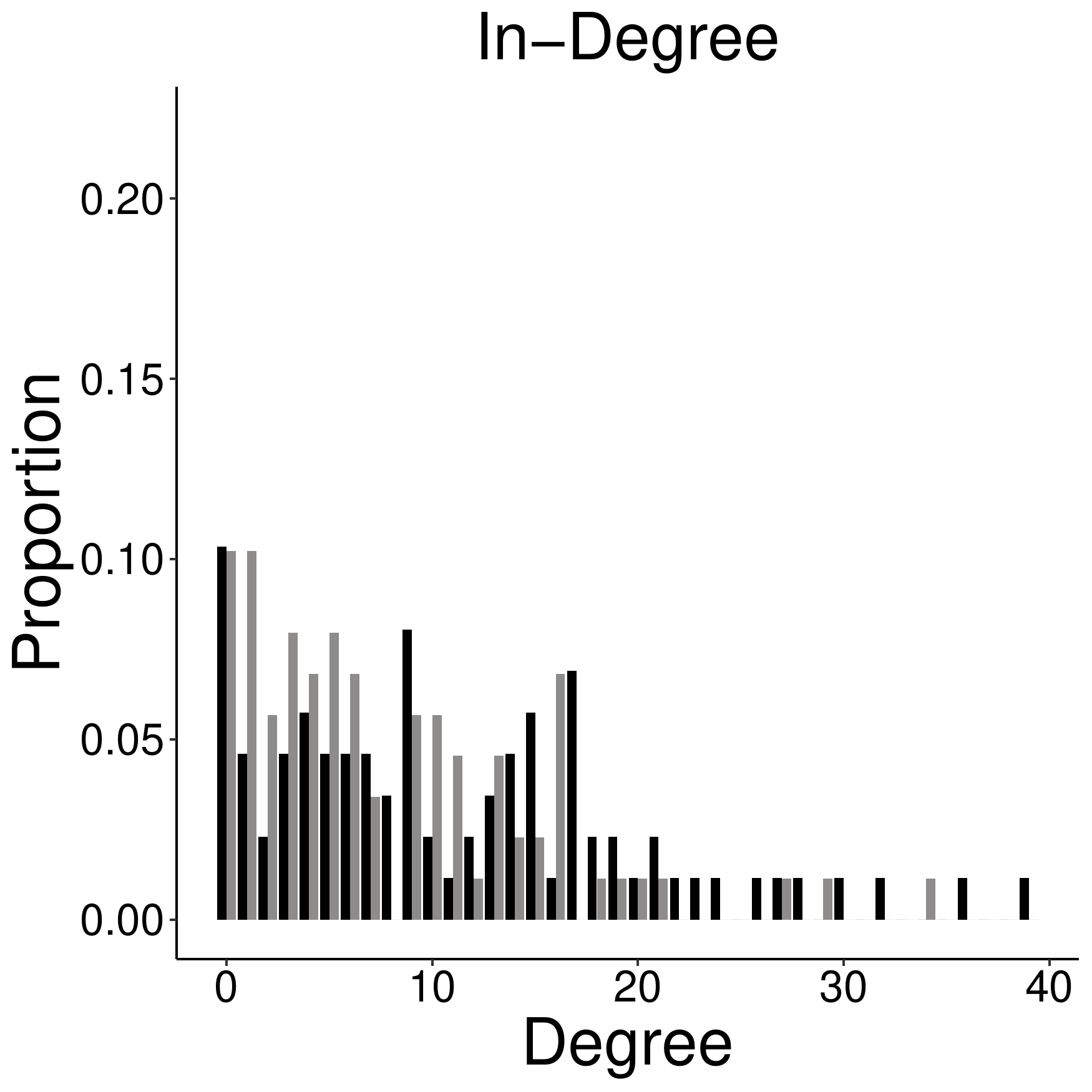}
		\includegraphics[trim={0cm 0cm 0cm 0cm},clip,width=0.45\textwidth]{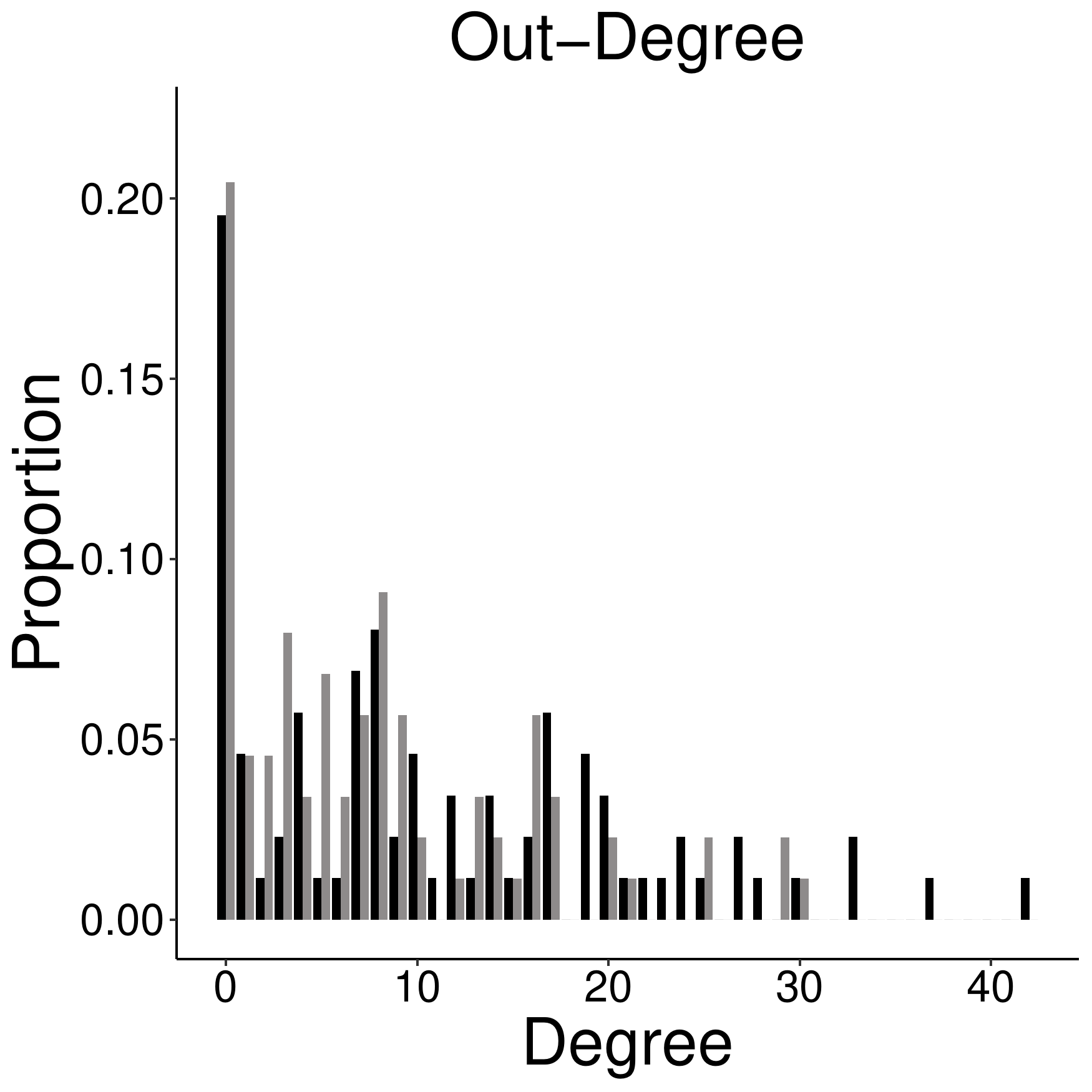}
		\caption{Email exchange network: Barplots indicating the distribution of the in- and out-degrees. Black bars indicate the values of period 1 and grey bars of period 2.}
		\label{fig:degree2}
	\end{figure}

	Figures \ref{fig:degree} and \ref{fig:degree2} depict the distributions of in- and out-degrees in the two networks. Building on in- and out-degree of all nodes, these distributions represent the relative frequency of all possible in- and out-degrees in the observed networks, which is calculated with the $\mathtt{igraph}$ package in \texttt{R} \citep{csardi2006}.
	
	In the arms trade network, a strongly asymmetric relation is revealed, indicating that about 70$\%$ of the countries do not export any weapons, while a small percentage of countries accounts for the major share of trade relations. The distribution of the in-degree is not that extreme but still we have roughly one third of all countries not importing at all. 
	
		The email exchange network shows a different structure. Here, many medium-sized in-degrees can be found and only roughly $10\%$ of all nodes have receiver no emails. For the out-degree, this number doubles (roughly 20\% have not sent emails). Further, the distribution of the out-degree is more skewed then the one for the in-degree.
	
\newpage
\clearpage

\section{Supplementary Material}
\subsection{Countries in the International Arms Trade Network}
	\FloatBarrier

\begin{table}[ht!]
\scriptsize 
\center
 \renewcommand{\arraystretch}{0.45}
\begin{tabular}{llllll}
  \hline
Country Name & ISO3 & Country Name & ISO3 & Country Name & ISO3 \\ 
  \hline
Afghanistan & AFG & Germany & DEU & Niger & NER \\ 
  Albania & ALB & Ghana & GHA & Nigeria & NGA \\ 
  Algeria & DZA & Greece & GRC & Norway & NOR \\ 
  Andorra & AND & Grenada & GRD & Oman & OMN \\ 
  Angola & AGO & Guatemala & GTM & Pakistan & PAK \\ 
  Antigua and Barbuda & ATG & Guinea & GIN & Palau & PLW \\ 
  Argentina & ARG & Guinea-Bissau & GNB & Panama & PAN \\ 
  Armenia & ARM & Guyana & GUY & Papua New Guinea & PNG \\ 
  Australia & AUS & Haiti & HTI & Paraguay & PRY \\ 
  Austria & AUT & Honduras & HND & Peru & PER \\ 
  Azerbaijan & AZE & Hungary & HUN & Philippines & PHL \\ 
  Bahamas & BHS & Iceland & ISL & Poland & POL \\ 
  Bahrain & BHR & India & IND & Portugal & PRT \\ 
  Bangladesh & BGD & Indonesia & IDN & Qatar & QAT \\ 
  Barbados & BRB & Iran & IRN & Romania & ROM \\ 
  Belarus & BLR & Iraq & IRQ & Russia & RUS \\ 
  Belgium & BEL & Ireland & IRL & Rwanda & RWA \\ 
  Belize & BLZ & Israel & ISR & Saint Kitts and Nevis & KNA \\ 
  Benin & BEN & Italy & ITA & Saint Lucia & LCA \\ 
  Bhutan & BTN & Jamaica & JAM & Saint Vincent and the Grenadines & VCT \\ 
  Bolivia & BOL & Japan & JPN & Samoa & WSM \\ 
  Botswana & BWA & Jordan & JOR & San Marino & SMR \\ 
  Brazil & BRA & Kazakhstan & KAZ & Sao Tome and Principe & STP \\ 
  Brunei Darussalam & BRN & Kenya & KEN & Saudi Arabia & SAU \\ 
  Bulgaria & BGR & South Korea & KOR & Senegal & SEN \\ 
  Burkina Faso & BFA & Kosovo & KOS & Serbia & YUG \\ 
  Burundi & BDI & Kuwait & KWT & Seychelles & SYC \\ 
  Cambodia & KHM & Kyrgyzstan & KGZ & Sierra Leone & SLE \\ 
  Cameroon & CMR & Laos & LAO & Singapore & SGP \\ 
  Canada & CAN & Latvia & LVA & Slovakia & SVK \\ 
  Cape Verde & CPV & Lebanon & LBN & Slovenia & SVN \\ 
  Central African Republic & CAF & Lesotho & LSO & Solomon Islands & SLB \\ 
  Chad & TCD & Liberia & LBR & South Africa & ZAF \\ 
  Chile & CHL & Libya & LBY & Spain & ESP \\ 
  China & CHN & Lithuania & LTU & Sri Lanka & LKA \\ 
  Colombia & COL & Luxembourg & LUX & Sudan & SDN \\ 
  Comoros & COM & Macedonia (FYROM) & MKD & Suriname & SUR \\ 
  DR Congo & ZAR & Madagascar & MDG & Swaziland & SWZ \\ 
  Congo & COG & Malawi & MWI & Sweden & SWE \\ 
  Costa Rica & CRI & Malaysia & MYS & Switzerland & CHE \\ 
  Cote dIvoire & CIV & Maldives & MDV & Tajikistan & TJK \\ 
  Croatia & HRV & Mali & MLI & Tanzania & TZA \\ 
  Cuba & CUB & Malta & MLT & Thailand & THA \\ 
  Cyprus & CYP & Marshall Islands & MHL & Timor-Leste & TMP \\ 
  Czech Republic & CZE & Mauritania & MRT & Togo & TGO \\ 
  Denmark & DNK & Mauritius & MUS & Trinidad and Tobago & TTO \\ 
  Dominica & DMA & Mexico & MEX & Tunisia & TUN \\ 
  Dominican Republic & DOM & Micronesia & FSM & Turkey & TUR \\ 
  Ecuador & ECU & Moldova & MDA & Turkmenistan & TKM \\ 
  Egypt & EGY & Mongolia & MNG & Uganda & UGA \\ 
  El Salvador & SLV & Montenegro & YUG & Ukraine & UKR \\ 
  Equatorial Guinea & GNQ & Morocco & MAR & United Arab Emirates & ARE \\ 
  Estonia & EST & Mozambique & MOZ & United Kingdom & GBR \\ 
  Ethiopia & ETH & Myanmar & MYM & United States & USA \\ 
  Fiji & FJI & Namibia & NAM & Uruguay & URY \\ 
  Finland & FIN & Nauru & NRU & Uzbekistan & UZB \\ 
  France & FRA & Nepal & NPL & Vanuatu & VUT \\ 
  Gabon & GAB & Netherlands & NLD & Viet Nam & VNM \\ 
  Gambia & GMB & New Zealand & NZL & Zambia & ZMB \\ 
  Georgia & GEO & Nicaragua & NIC & Zimbabwe & ZWE \\ 
   \hline
\end{tabular}
\caption{Countries included in the analysis of the international trade network with the ISO3 codes, that are used in the graphical representations of the network. }
\end{table}

	\FloatBarrier

		\subsection{Simulation-based Goodness-of-fit in (S)TERGMs}	\label{sec:simgof}
\begin{figure}[ht!]\centering
	\includegraphics[trim={0cm 0cm 0cm 0cm},clip,width=0.89\textwidth]{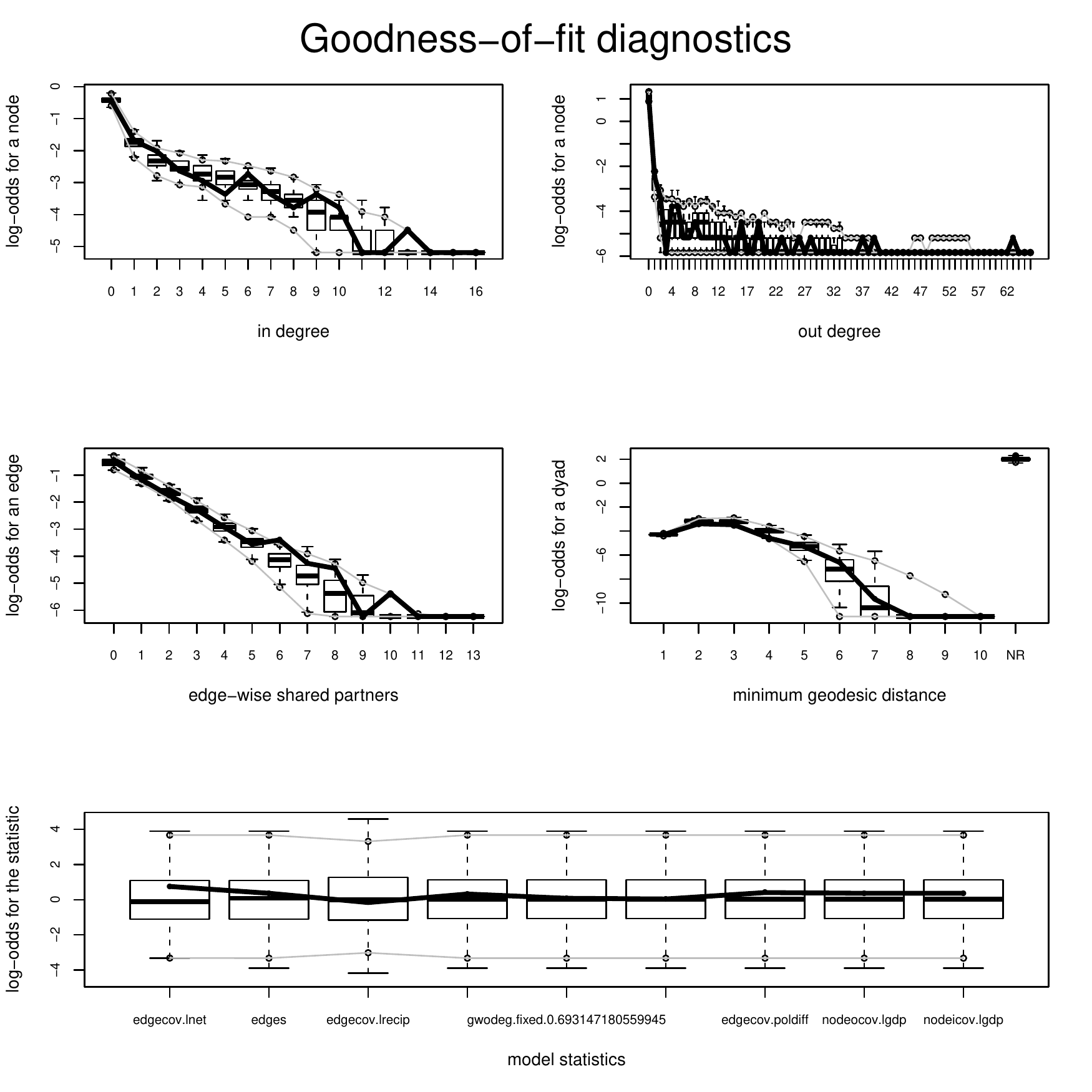}
	\caption{Arms trade network: Simulation-based goodnes-of-fit diagnostics in the TERGM. Boxplots give the evaluations of the respective network characteristics at the simulated networks and the solid line gives the actual values from the observed network.
		First four panels give the log-odds of a node for different in-degrees (top left), out-degrees (top right), edge-wise shared partners (middle, left) and minimum geodesic distance (middle right). All included rescaled network statistics on the bottom panel.}
	\label{fig:gof.tergm}
\end{figure}

\begin{figure}[ht!]\centering
		\includegraphics[page=1,trim={0cm 0cm 0cm 0cm},clip,width=0.89\textwidth]{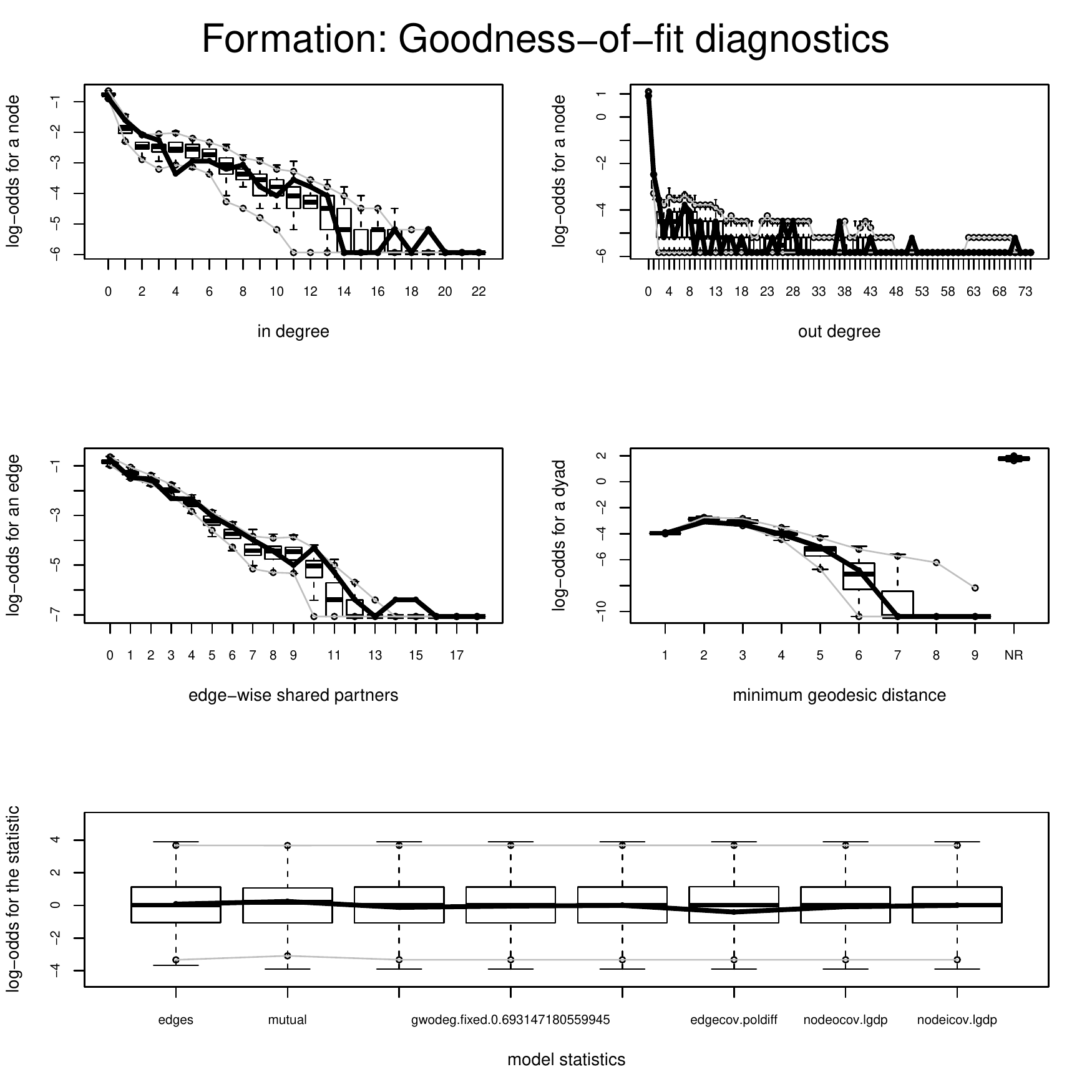}
	\caption{Arms trade network: Simulation-based goodnes-of-fit diagnostics in the STERGM for the formation model. Boxplots give the evaluations of the respective network characteristics at the simulated networks and the solid line gives the actual values from the observed network.
	First four panels give the log-odds of a node for different in-degrees (top left), out-degrees (top right), edge-wise shared partners (middle, left) and minimum geodesic distance (middle right). All included rescaled network statistics on the bottom panel.}
		\label{fig:gof.stergmf}
	\end{figure}

\begin{figure}[ht!]\centering
	\includegraphics[page=2,trim={0cm 0cm 0cm 0cm},clip,width=0.89\textwidth]{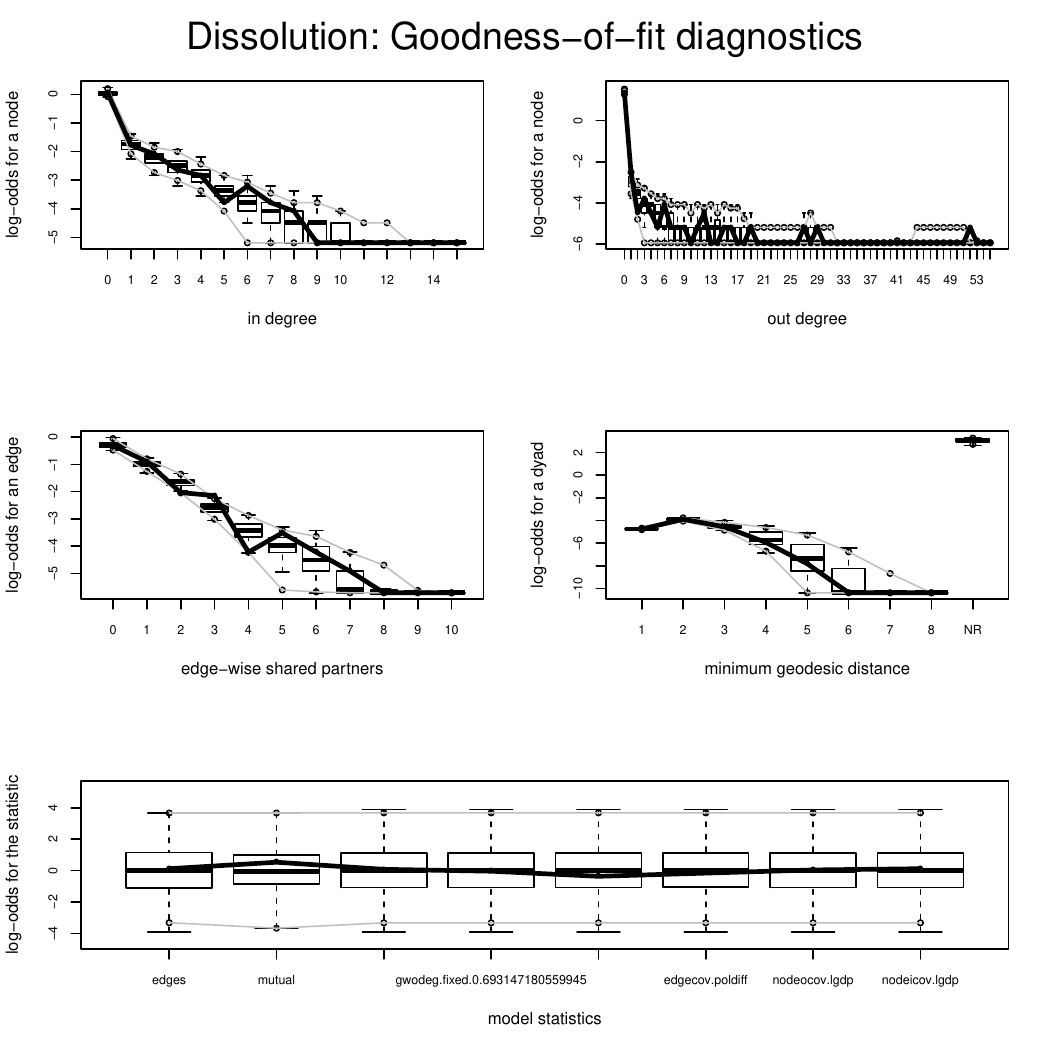}
	\caption{Arms trade network: Simulation-based Goodnes-of-fit diagnostics in the STERGM for the dissolution model. Boxplots give the evaluations of the respective network characteristics at the simulated networks and the solid line gives the actual values from the observed network.
	First four panels give the log-odds of a node for different in-degrees (top left), out-degrees (top right), edge-wise shared partners (middle, left) and minimum geodesic distance (between them, middle right). All included rescaled network statistics on the bottom panel.}
	\label{fig:gof.stergmd}
\end{figure}

\begin{figure}[ht!]\centering
			\includegraphics[trim={0cm 0cm 0cm 0cm},clip,width=0.89\textwidth]{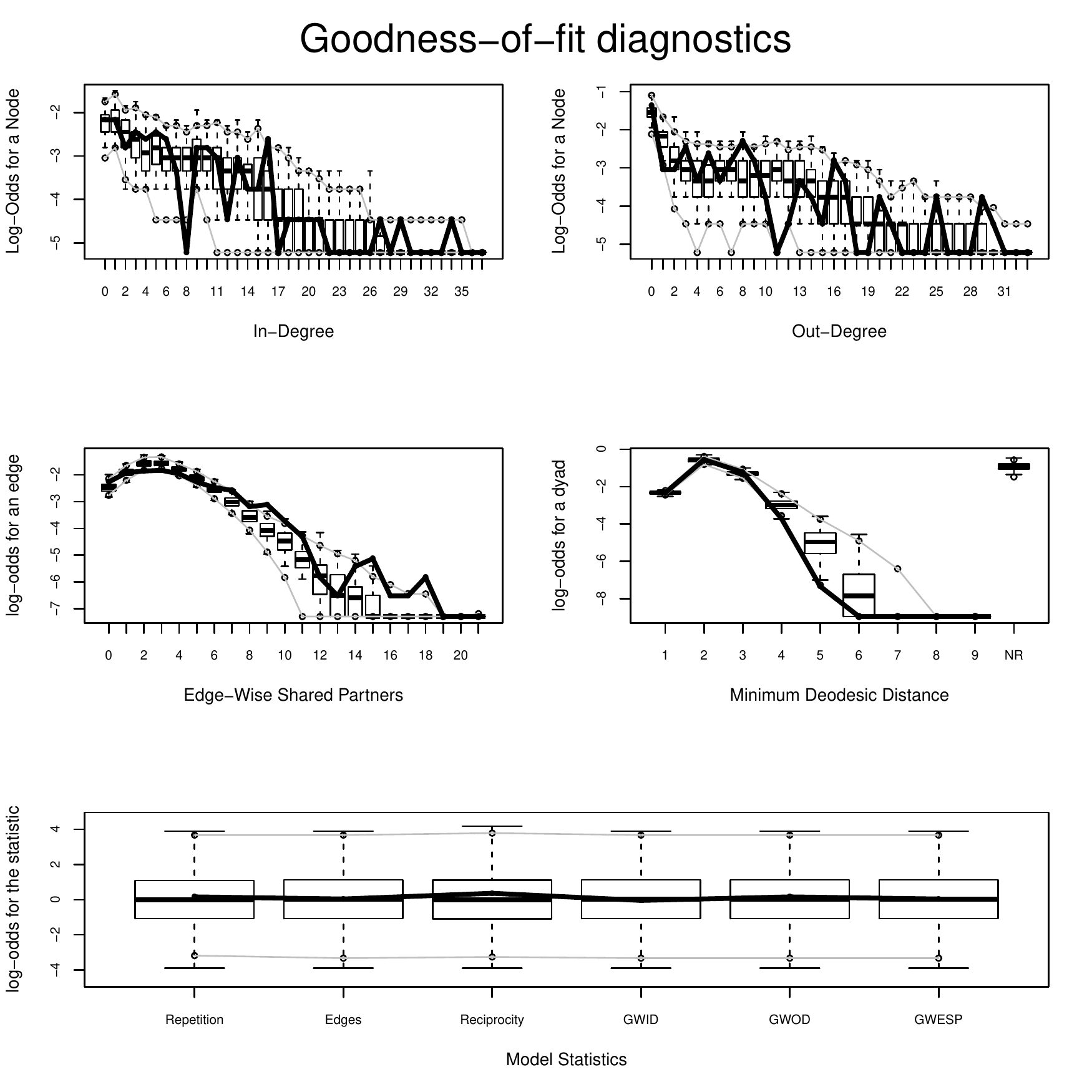}
			\caption{Email exchange network: Simulation-based Goodnes-of-fit diagnostics in the TERGM. Boxplots give the evaluations of the respective network characteristics at the simulated networks and the solid line gives the actual values from the observed network.
				First four panels give the log-odds of a node for different in-degrees (top left), out-degrees (top right), edge-wise shared partners (middle, left) and minimum geodesic distance (middle right). All included rescaled network statistics on the bottom panel.}
			\label{fig:gof.email.tergm}
		\end{figure}

\begin{figure}[ht!]\centering
			\includegraphics[page=1,trim={0cm 0cm 0cm 0cm},clip,width=0.89\textwidth]{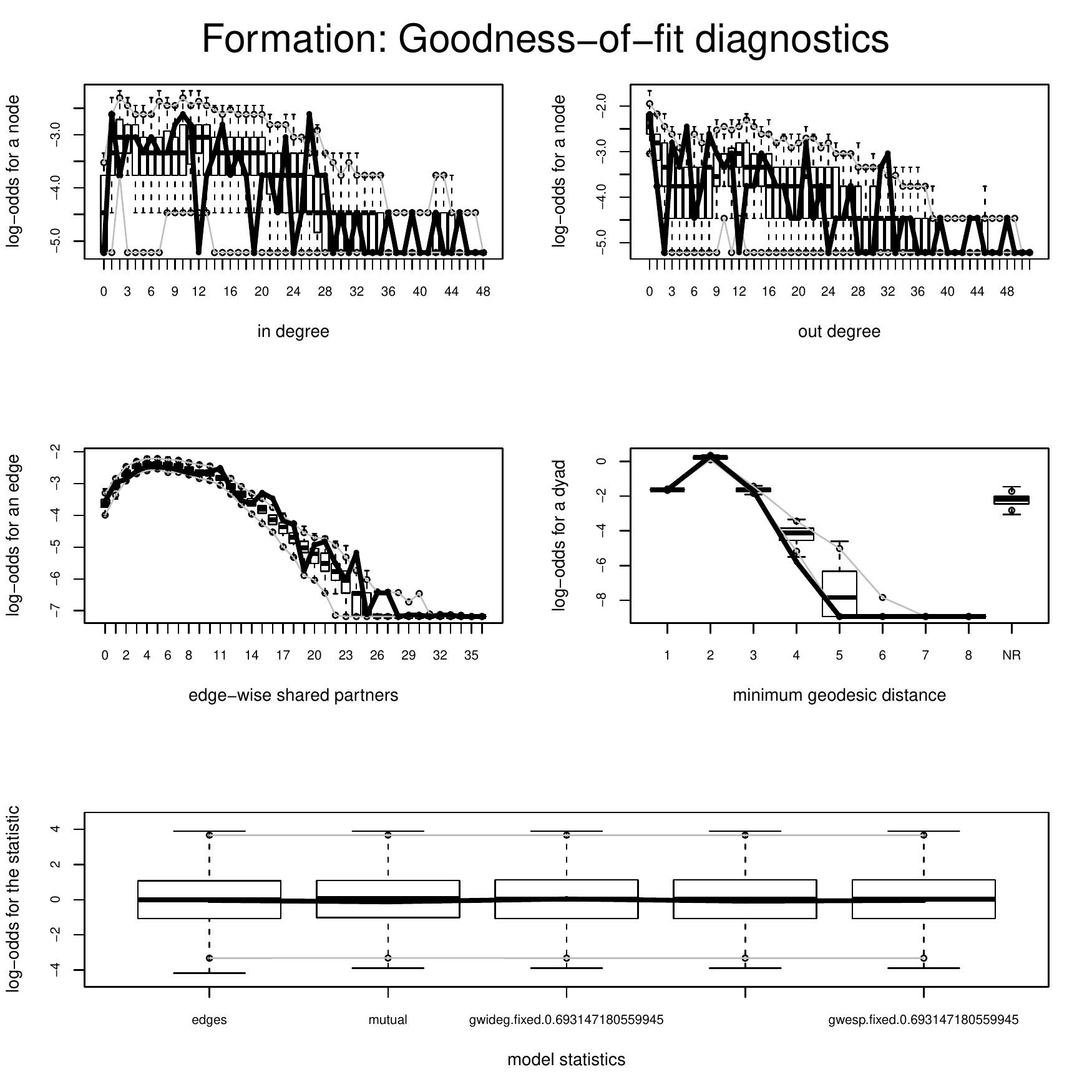}
			\caption{Email exchange network: Simulation-based Goodnes-of-fit diagnostics in the STERGM for the formation model. Boxplots give the evaluations of the respective network characteristics at the simulated networks and the solid line gives the actual values from the observed network.
				First four panels give the log-odds of a node for different in-degrees (top left), out-degrees (top right), edge-wise shared partners (middle, left) and minimum geodesic distance (middle right). All included rescaled network statistics on the bottom panel.}
			\label{fig:gof.email.stergmf}
		\end{figure}

\begin{figure}[ht!]\centering
			\includegraphics[page=2,trim={0cm 0cm 0cm 0cm},clip,width=0.89\textwidth]{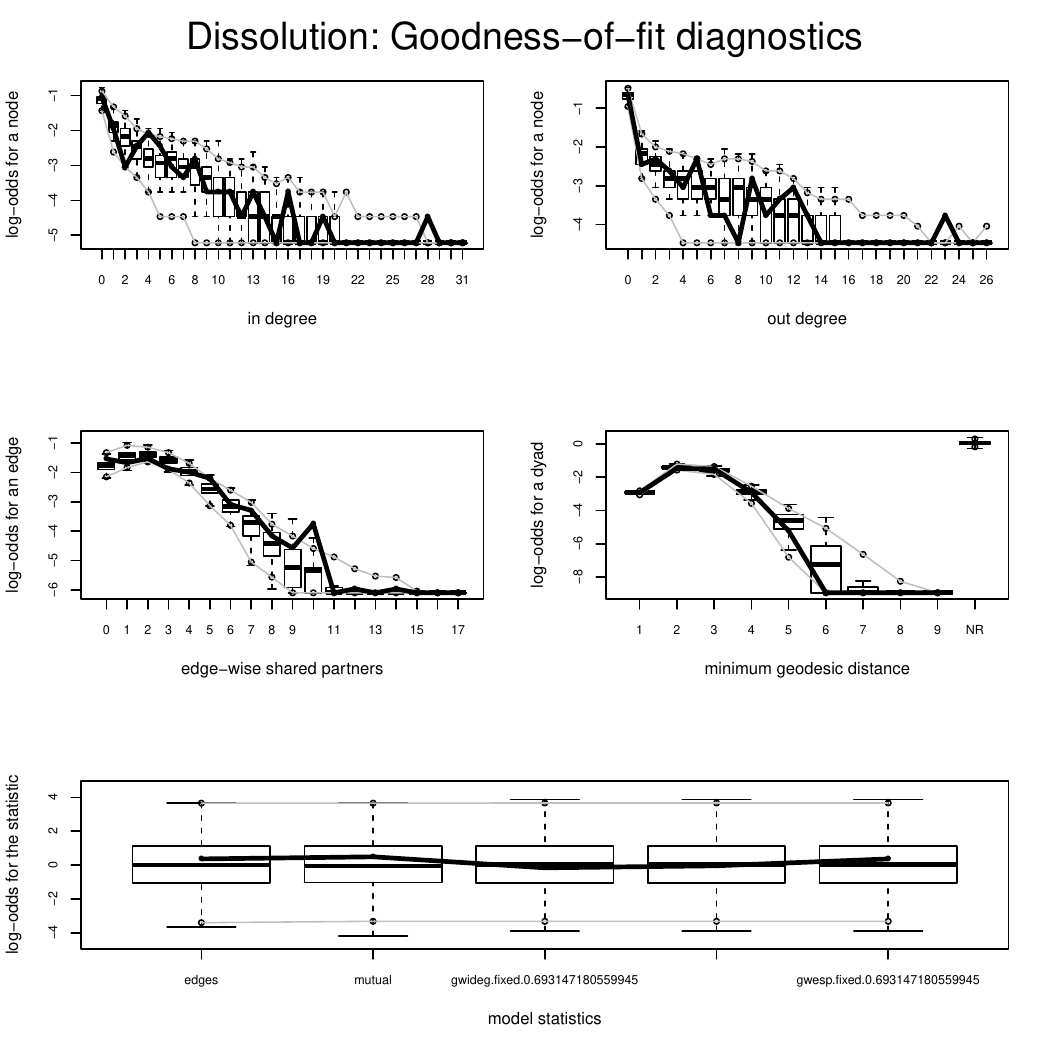}
			\caption{Email exchange network: Simulation-based Goodnes-of-fit diagnostics in the STERGM for the dissolution model. Boxplots give the evaluations of the respective network characteristics at the simulated networks and the solid line gives the actual values from the observed network.
				First four panels give the log-odds of a node for different in-degrees (top left), out-degrees (top right), edge-wise shared partners (middle, left) and minimum geodesic distance (between them, middle right). All included rescaled network statistics on the bottom panel.}
			\label{fig:gof.email.stergmd}
		\end{figure}

			In Figures \ref{fig:gof.tergm} and \ref{fig:gof.email.tergm} we show simulation-based godness-of-fit (GOF) diagnostics for the the TERGM model and in Figures \ref{fig:gof.stergmf}, \ref{fig:gof.stergmd}, \ref{fig:gof.email.stergmf} and \ref{fig:gof.email.stergmd} for the STERGM in the formation and dissolution model, respectively. The figures are created by the \texttt{R} package \texttt{ergm} (version 3.10.4) and follow the approach of \citet{Hunter2008a}. In all three models, the fitted model is used in order to simulate 100 new networks. Based on these, different network characteristics are computed and visualized in boxplots.
			
			 The standard characteristics used are the complete distributions of the in-degree, out-degree, edge-wise shared partners and minimum geodesic distance (i.e.\ number of node pairs with shortest path of length $k$ between them). The solid black line indicates the measurements of these characteristic in the observed network. These statistics show whether measures like GWID, GWOD and GWESP are sufficient to reproduce global network patterns. Because many shares are rather small, we visualize the simulated and observed measures on a log-odds scale. 
			 
			 On the bottom of the figures it is shown how well the actual network statistics are reproduced. Note, that both models compare different things as the TERGM is evaluated at $y_t$ while the STERGM regards $y^+$ and $y^-$. Overall, all plots indicate a satisfying fit of the respective models. 
			  
	\newpage 
	\subsubsection{Data Set 1: International Arms Trade}
	\FloatBarrier
		\FloatBarrier

	\FloatBarrier

		\FloatBarrier
		\newpage 
	\FloatBarrier

		\subsubsection{Data Set 2: European Research Institution Email Correspondence }
		\FloatBarrier
			\FloatBarrier

			\FloatBarrier

			\FloatBarrier

	\newpage
	
	\subsection{ROC-based Goodness-of-fit}	\label{sec:simgof}
	\subsubsection{Data Set 1: International Arms Trade}

As already stated in Section \ref{sec:goodness-pf-fit-discrete}  of the main paper techniques for assessing the fit of a probabilistic classification can be used when working with binary network data. In the case of observations at discrete time points this allows an informal comparison of the models proposed in Section \ref{sec:discrete}  and \ref{Time-Clustered Observations}. 

\begin{figure}[ht!]\centering
		\includegraphics[page=1,trim={0cm 0cm 0cm 0cm},clip,width=0.48\textwidth]{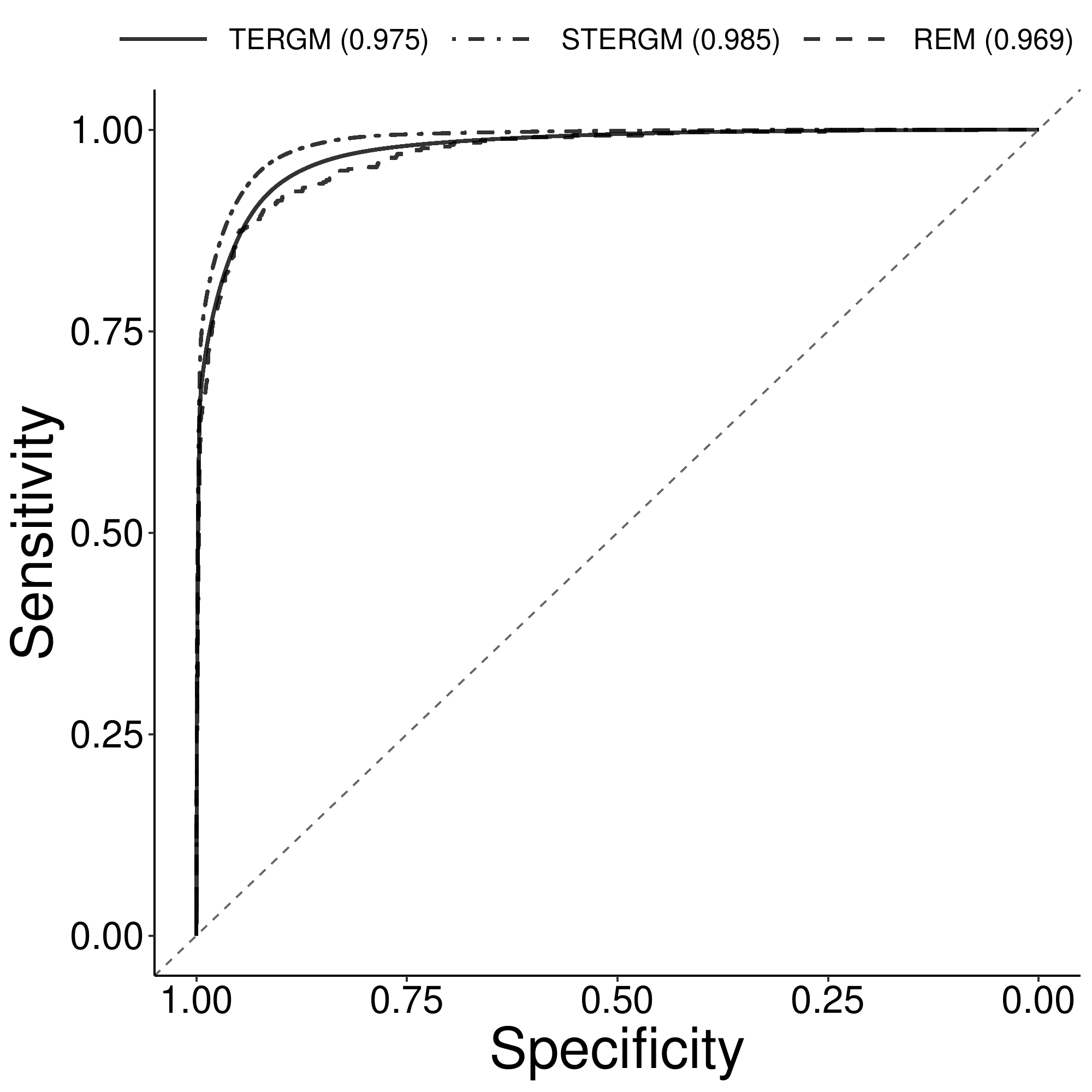}
		\includegraphics[page=1,trim={0cm 0cm 0cm 0cm},clip,width=0.48\textwidth]{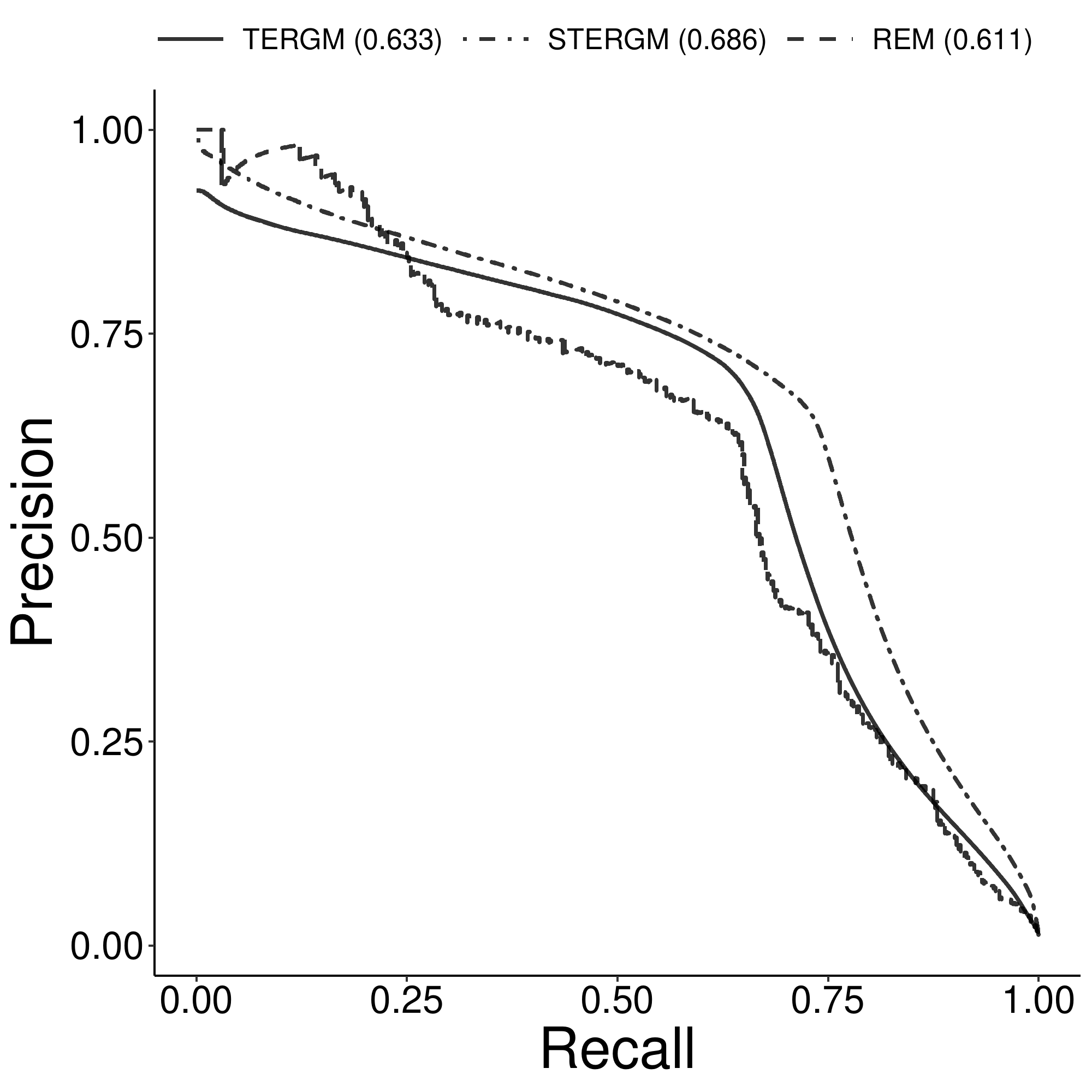}
	\caption{Arms trade network: ROC and PR curves from the TERGM (dotted line), STERGM (dotdashed line), and REM (solid line).The AUC values of the respective curves are indicated in brackets. }
		\label{fig:gof.comp}
	\end{figure}

In the case of the TERGM and STERGM the application of the ROC- and PR-curve follows from the conditional probability of observing a specific tie (see equation (2.5) of \citealp{hunter2006}). For the REM we predict the intensities of all possible events given the information of $t-1$ and use this value as a score in the calculation of the ROC curve. While the latter approach is non-standard and can only be applied to REMs that regard durable ties, it enables a direct comparison between the models as shown in Figure \ref{fig:gof.comp}. The results of the ROC curve indicate a generally good fit of all models. In the STERGM more parameters are estimated, which seems to lead to a slightly bigger area under curve (AUC) values as compared to the REM and TERGM. Similar to the conclusions from the ROC curve, the PR curve favors the TERGM and STERGM over the REM. 

		\newpage 

		\subsubsection{Data Set 2: European Research Institution Email Correspondence }

\begin{figure}[ht!]\centering
		\includegraphics[page=1,trim={0cm 0cm 0cm 0cm},clip,width=0.48\textwidth]{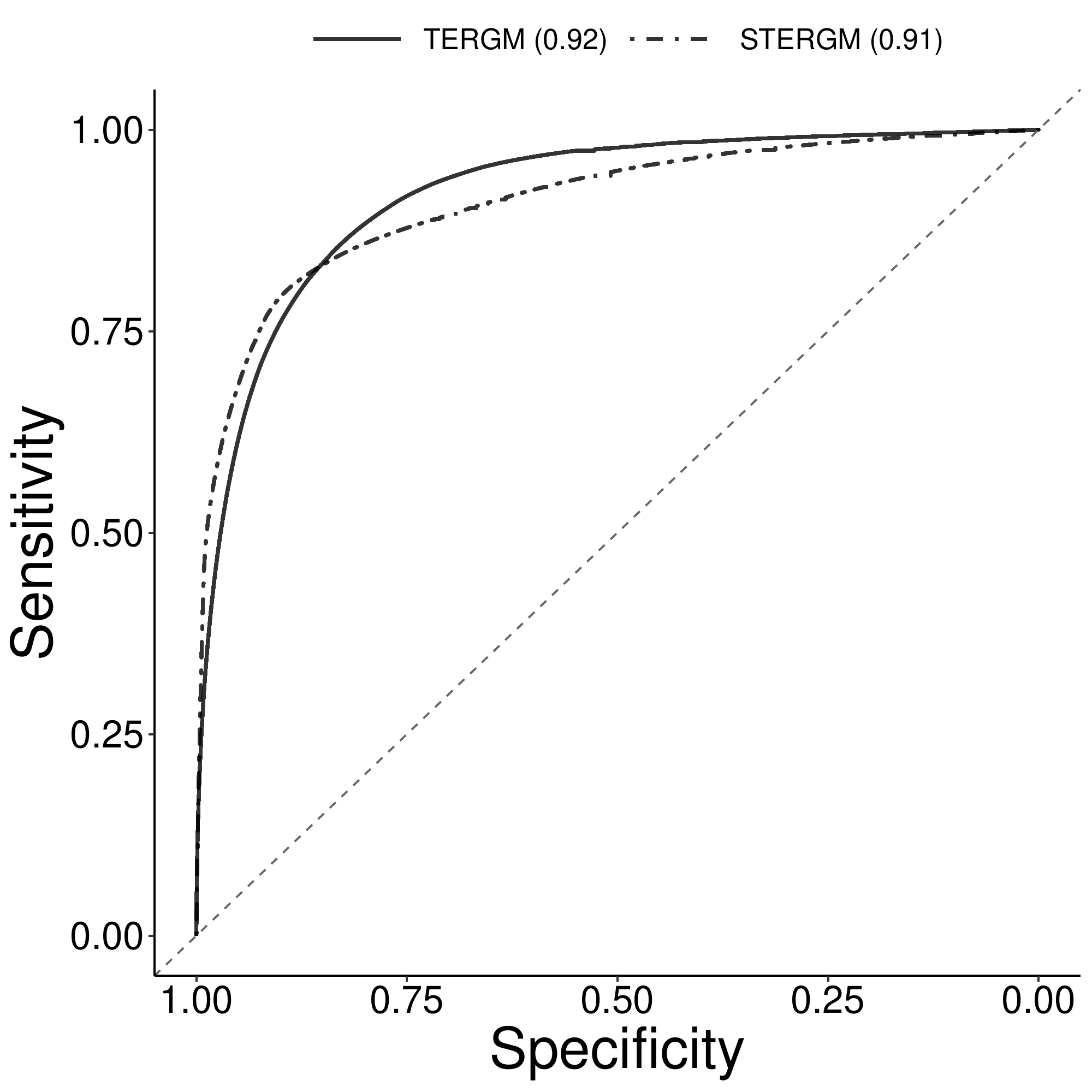}
		\includegraphics[page=1,trim={0cm 0cm 0cm 0cm},clip,width=0.48\textwidth]{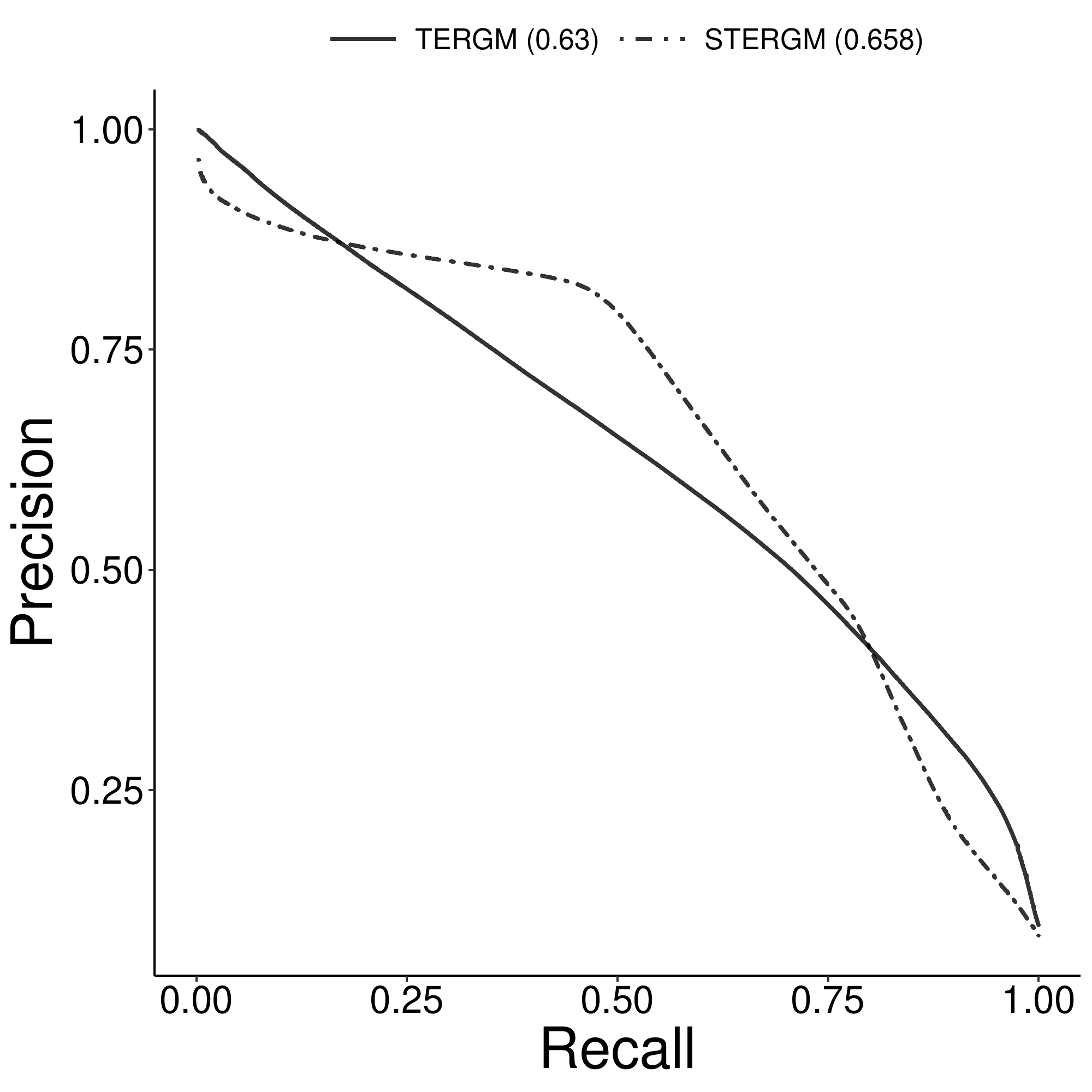}
	\caption{Email exchange network: ROC and PR curves from the TERGM (dotted line), STERGM (dotdashed line), and REM (solid line). The AUC values of the respective curves are indicated in brackets. }
		\label{fig:gof.comp2}
	\end{figure}

The second data set regards reoccurring events in the REM, which are aggregated for the analysis of the TERGM and STERGM. Therefore, the ROC and PR curve are only available for the TERGM and STERGM. The results are depicted in Figure \ref{fig:gof.comp2}. For this data set, the ROC curves favor the TERGM. Yet, when emphasis is put on finding the true positives, the PR curve detects a better model fit of the STERGM.  

\begin{figure}[ht!]\centering
 	
 	 	\begin{subfigure}[b]{0.49\linewidth}
 \centering \includegraphics[page=1,trim={0cm 0cm 0cm 0cm},clip,width=1\textwidth]{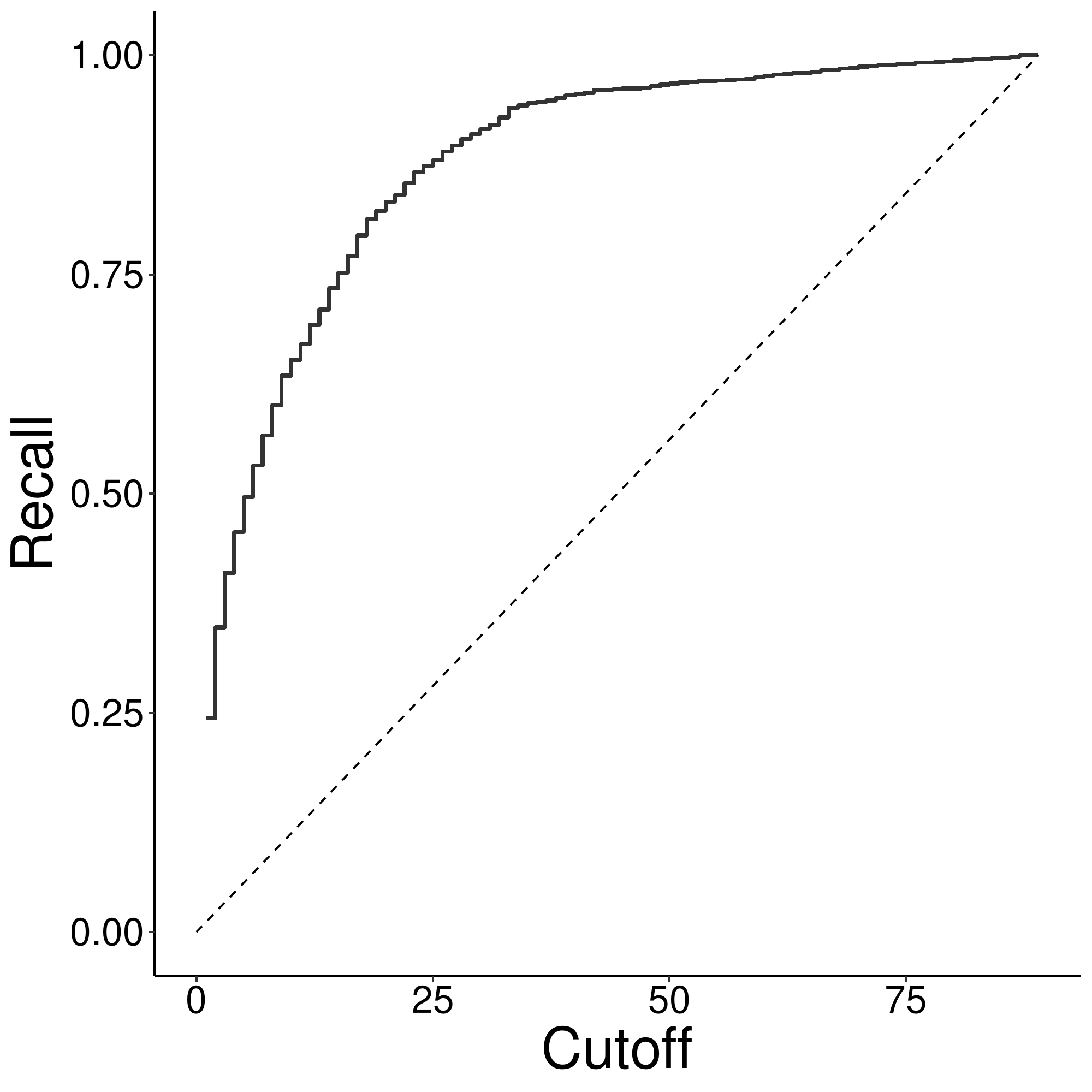}
  \caption{\centering \label{1} }
 \end{subfigure}
 \begin{subfigure}[b]{0.49\linewidth}
 \centering \includegraphics[page=1,trim={0cm 0cm 0cm 0cm},clip,width=1\textwidth]{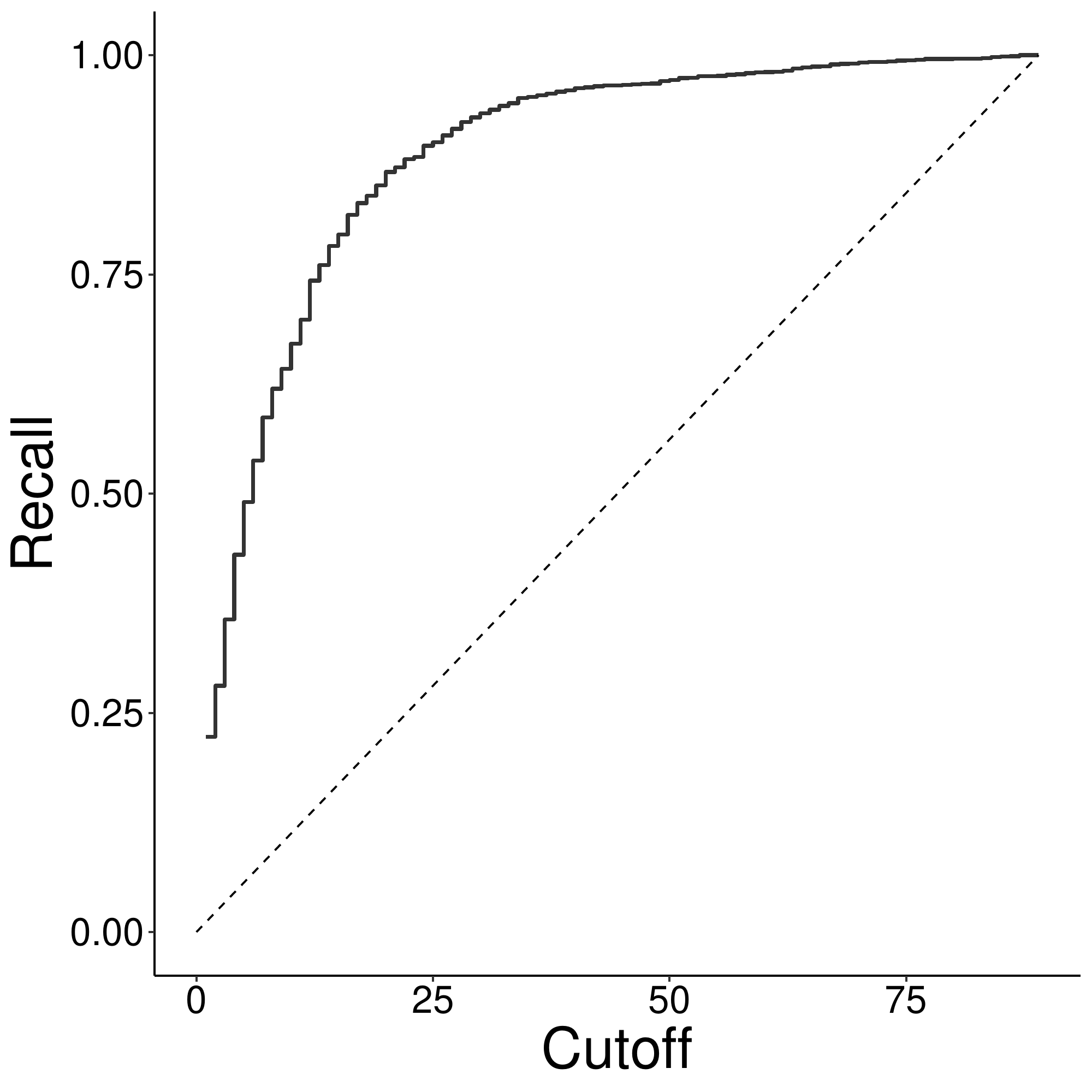}
  \caption{\centering \label{2} }
 \end{subfigure}
 \begin{subfigure}[b]{0.49\linewidth}
 \centering \includegraphics[page=1,trim={0cm 0cm 0cm 0cm},clip,width=1\textwidth]{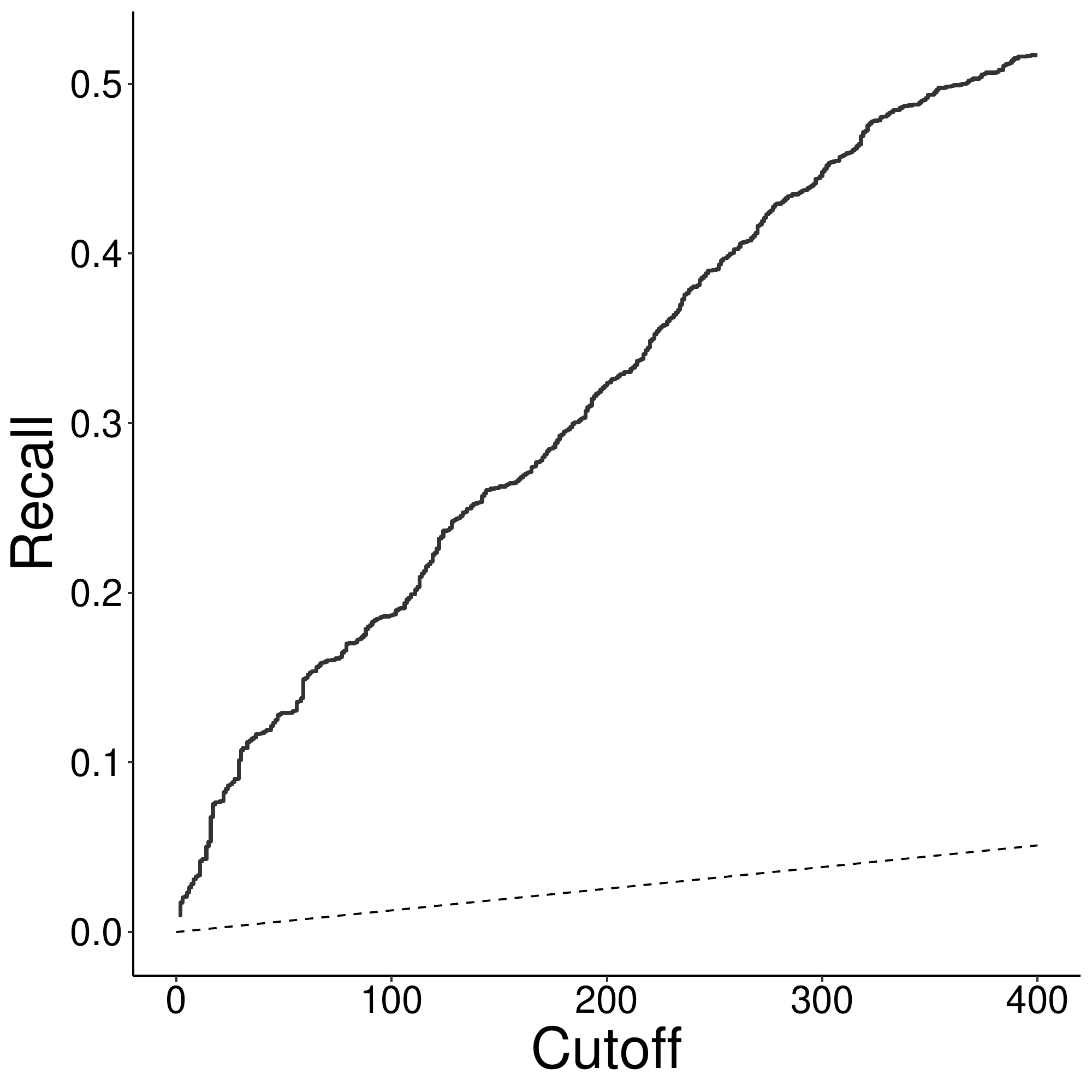}
  \caption{\centering \label{3} }
 \end{subfigure}
	\caption{Email exchange network: Recall Curves of REM regarding the next sender (a), receiver (b), and event (c). The dotted line indicates the measure under random guessing of the next sender, receiver, or event.}
		\label{fig:gof.recall}
	\end{figure}

As explained in Section 4.4 one option to asses the goodness-of-fit of REMs is the recall measure as proposed by \citet{vu2011_2}. We apply the measure in three different situations that may be of interest when measuring the predictive performance of relational event models: predict the next tie, next sender, and next receiver. The worst case scenario in terms of predictions of a model would be random guessing of the next sender, receiver, or event, the resulting recall rates are indicated by the dotted lines. The results in Figure \ref{fig:gof.recall} exhibit a good predictive performance of the REM, i.e. in about 75$\%$ of the events the right sender and receiver is among the 25 most likely senders and receivers. 
	\FloatBarrier
	\subsection{Application with multiple time points}
In the main article, we fitted a TERGM as well as a STERGM to two time points, called \textsl{period 1} and \textsl{period 2}. However, it is possible to fit these models to multiple transitions. In order to do so, we took the first two years of the email exchange network and aggregated the deciles into 10 binary networks.  Using again a first order Markov assumption and conditioning on the first network, this allows to fit a TERGM as well as a STERGM to the remaining nine networks. For comparison we additionally fit a REM to the data set. The estimation is done using the function \texttt{mtergm} 	from the package \texttt{btergm} (version 3.6.1) (\citealp{leifeld2017}) that implements MCMC-based maximum likelihood. The STERGM is fitted again using the package \texttt{tergm} (version 3.5.2). 

The corresponding results can be found in Table \ref{table:coefficients} in column 1 to 3. Note that the parameter estimates still refer to the transition from $t-1$ to $t$ and can interpreted in the same way as in the main article. In that regard note, that it is now assumed that the coefficients stay constant with time. Possible approaches to relax this assumption were given in the main article. The estimates of the REM (column 4) are consistent with the (S)TERGM but slightly differ to the main article, since now we condition only on the first out of 10 \textsl{periods} and hence model more events. 

\begin{table}
\begin{center}
\begin{tabular}{l c c c | c r}
	\hline
			& TERGM &  \multicolumn{2}{c|}{STERGM} & REM \\ 
			&  & Formation & Dissolution  & \\ 
			\hline \hline 
Repetition       & $1.986^{***}$  & $-$  & $-$  & $2.354^{***}$& \\
                 & $(0.058)$      &     $-$   &   $-$     & $(0.048)$   & $-$ \\
Edges            & $-3.435^{***}$ & $-4.485^{***}$ & $-0.769^{***}$ & $-$ &\\
                 & $(0.059)$      & $(0.073)$      & $(0.107)$      &  $-$  & \\
Reciprocity      & $1.090^{***}$  & $2.755^{***}$  & $1.761^{***}$  & $1.699^{***}$ &\\
                 & $(0.069)$      & $(0.079)$      & $(0.113)$      & $(0.043)$  &   \\
In-Degree (GWID)  & $-1.239^{***}$ & $-0.870^{***}$ & $-0.307^{*}$   & $0.007^{***}$& In-Degree Receiver \\
                 & $(0.099)$      & $(0.156)$      & $(0.152)$      & $(0.002)$     \\
Out-Degree (GWOD) & $-1.574^{***}$ & $-1.757^{***}$ & $-0.056$       & $0.009^{***}$& Out-Degree Sender \\
                 & $(0.097)$      & $(0.155)$      & $(0.159)$      & $(0.001)$     \\
GWESP            & $0.497^{***}$  & $0.508^{***}$  & $0.124^{**}$   & $0.199^{***}$& Transitivity \\
                 & $(0.028)$      & $(0.029)$      & $(0.043)$      & $(0.011)$     \\
\hline

\end{tabular}
\caption{Email exchange network: Comparison of parameters obtained from the  TERGM (first column) and STERGM (Formation in the second column, Dissolution in the third column). Standard errors in brackets and stars according to $p$-values smaller than $0.001$ ($^{***}$), $0.05$ ($^{**}$) and $0.1$ ($^{*}$). Decay parameter of the geometrically weighted statistics is set to $\log(2)$.}
\label{table:coefficients}
\end{center}
\end{table}

\end{document}